\documentclass[conference]{IEEEtran}
\ifCLASSINFOpdf
\else
\fi

\usepackage[hidelinks]{hyperref}
\usepackage{url}
\usepackage{xcolor}
\usepackage{xspace}
\usepackage[binary-units=true,per-mode=symbol,mode=text,detect-all]{siunitx}
\DeclareSIUnit{\Bit}{\text{Bit}}
\DeclareSIUnit{\million}{Mio.}
\DeclareSIUnit{\billion}{Bil.}

\usepackage{booktabs}
\usepackage{bm}
\usepackage{marvosym}
\usepackage{amssymb}
\usepackage{cite}

\newif\ifdraft

\drafttrue
\draftfalse

\ifdraft
\newcommand{\todo}[1]{{\color{red}{\textbf{TODO:} #1}}}

\newcommand{\new}[1]{{\color{green}{#1}}}
\newcommand{\jp}[1]{{\color{magenta}{#1}}}
\newcommand{\eb}[1]{{\color{blue}{#1}}}
\newcommand{\md}[1]{{\color{olive}{#1}}}
\newcommand{\mh}[1]{{\color{green}{#1}}}

\else
\newcommand{\todo}[1]{}

\newcommand{\new}[1]{{#1}}
\newcommand{\jp}[1]{{}}
\newcommand{\eb}[1]{{}}
\newcommand{\md}[1]{{}}
\newcommand{\mh}[1]{{}}
\fi

\newcommand{\ie}{i.e.,\xspace}
\newcommand{\eg}{e.g.,\xspace}

\newcommand{\internalstep}[2][]{\raisebox{.5pt}{\textcircled{\raisebox{-.5pt}{\small#2\raisebox{0.5pt}{\footnotesize #1}}}}}

\newcommand{\step}[2][]{Match~\internalstep[#2]{#1}}

\usepackage{tikz}

\newcommand*\rectangle[1]{\tikz[baseline=(char.base)]{
            \node[shape=rectangle,draw,inner sep=1.5pt] (char) {#1};}}

\newcommand{\stephighlight}[1]{\rectangle{\textbf{#1}}}

\makeatletter
\newcommand{\setpaper}[3]{%
  \phantomsection
  #1 (#2)\def\@currentlabel{\unexpanded{\textbf{#2}}}\label{#3}%
}
\makeatother

\begin{document}
\bstctlcite{IEEEexample:BSTcontrol}
%
\title{Collaboration is not Evil: A Systematic Look\\ at Security Research for Industrial Use}

\author{
  \IEEEauthorblockN{
    Jan Pennekamp\IEEEauthorrefmark{1},
    Erik Buchholz\IEEEauthorrefmark{1},
    Markus Dahlmanns\IEEEauthorrefmark{1},
    Ike Kunze\IEEEauthorrefmark{1},
    Stefan Braun\IEEEauthorrefmark{2},
    Eric Wagner\IEEEauthorrefmark{3}\IEEEauthorrefmark{1},
    \\
    Matthias Brockmann\IEEEauthorrefmark{4},
    Klaus Wehrle\IEEEauthorrefmark{1},
    Martin Henze\IEEEauthorrefmark{5}\IEEEauthorrefmark{3}
  }

  \IEEEauthorblockA{
    \IEEEauthorrefmark{1}\textit{Communication and Distributed Systems},
    \textit{RWTH Aachen University}, Germany $\cdot$
    \{lastname\}@comsys.rwth-aachen.de
  }
  \IEEEauthorblockA{
    \IEEEauthorrefmark{2}\textit{Databases and Information Systems},
    \textit{RWTH Aachen University}, Germany $\cdot$
    braun@dbis.rwth-aachen.de
  }
  \IEEEauthorblockA{
    \IEEEauthorrefmark{3}\textit{Cyber Analysis \& Defense},
    \textit{Fraunhofer FKIE}, Germany $\cdot$
    \{firstname.lastname\}@fkie.fraunhofer.de
  }
  \IEEEauthorblockA{
    \IEEEauthorrefmark{4}\textit{Machine Tools and Production Engineering},
    \textit{RWTH Aachen University}, Germany $\cdot$
    m.brockmann@wzl.rwth-aachen.de
  }
  \IEEEauthorblockA{
    \IEEEauthorrefmark{5}\textit{Security and Privacy in Industrial Cooperation},
    \textit{RWTH Aachen University}, Germany $\cdot$
    henze@cs.rwth-aachen.de
  }
}


%


\IEEEoverridecommandlockouts
\makeatletter\def\@IEEEpubidpullup{6.5\baselineskip}\makeatother
\IEEEpubid{\parbox{\columnwidth}{
    {\fontsize{8.5}{8.5}\selectfont Learning from Authoritative Security Experiment Results (LASER) 2020 \\
    8 December 2020, Virtual \\
    \url{https://dx.doi.org/10.14722/laser-acsac.2020.23088} \\
    \url{www.acsac.org}}
}
\hspace{\columnsep}\makebox[\columnwidth]{}}

\maketitle

\begin{abstract}
\new{%
Following the recent Internet of Things-induced trends on digitization in general, industrial applications will further evolve as well.
With a focus on the domains of manufacturing and production, the Internet of Production pursues the vision of a digitized, globally interconnected, yet secure environment by establishing a distributed knowledge base.}\\
\noindent\underline{\smash{Background.}}
As part of our collaborative research of advancing the scope of industrial applications through cybersecurity and privacy, we identified a set of common challenges and pitfalls that surface in such \new{applied} interdisciplinary collaborations.\\
\noindent\underline{\smash{Aim.}}
Our goal with this paper is to support researchers in the emerging field of cybersecurity in industrial settings by formalizing our experiences as reference for other research efforts\new{, in industry and academia alike}.\\
\noindent\underline{\smash{Method.}}
Based on our experience, we derived a process cycle of performing such interdisciplinary research, from the initial idea to the eventual dissemination and paper writing.
This presented methodology strives to successfully bootstrap further research and to encourage further work in this emerging area.\\
\noindent\underline{\smash{Results.}}
\new{Apart from our newly proposed} process cycle, we report on our experiences and conduct a case study applying this methodology, raising awareness for challenges in cybersecurity research for industrial applications.
\new{%
We further detail the interplay between our process cycle and the data lifecycle in applied research data management.}
\new{Finally,} we augment our discussion with an industrial as well as an academic view on this research area and highlight that both areas still have to overcome significant challenges to sustainably and securely advance industrial applications.\\
\noindent\underline{\smash{Conclusions.}}
With our \new{proposed} process cycle for interdisciplinary research in the intersection of cybersecurity and industrial application, we provide a foundation for further research.
We look forward to promising research \new{initiatives}, projects, and directions that emerge based on our \new{methodological} work.
\end{abstract}


%

\section{Introduction}
\label{sec:introduction}

Under the umbrella of terms such as cyber-physical systems (CPS) \cite{Henzeetal2017Network}, the Internet of Production (IoP)~\cite{Pennekampetal2019Towards,Brauneretal2022A}, and the Industrial Internet of Things (IIoT)~\cite{Serroretal2021Challenges}, we observe an accelerating trend for digitization across all industrial sectors.
Within the scope of this digitization, cybersecurity research is not only indispensable to secure industrial networks and cyber-physical systems~\cite{Henzeetal2017Network}, but it also provides the opportunity to realize novel forms of industrial applications and collaboration:
By applying methods and tools developed by the security and privacy community to industrial use cases, we can provide functionality that was previously considered impossible due to prevalent confidentiality and privacy concerns.

As we have presented with our previous research efforts, recent advances in cybersecurity and privacy research allow the realization of secure multi-hop accountability in (physical) supply chains by cleverly combining tamper-resistance, which is provided by blockchains, with attribute-based encryption~\cite{Pennekampetal2020Private,Baderetal2021Blockchain-Based}.
Likewise, transparent end-to-end encryption allows companies to securely realize many-to-many communication on top of well-established communication protocols, which already serve as the foundation for various industrial CPS scenarios~\cite{Dahlmannsetal2021Transparent}.
Finally, different cryptographic building blocks interplay to lay the foundation to then create a privacy-preserving production process parameter exchange~\cite{Pennekampetal2020Privacy-Preserving} and real-world applicable company benchmarking~\cite{Pennekampetal2020Revisiting}.

Despite these promising forays, to date, only a few approaches \new{(\eg~\cite{Ballotetal2014The,Xu2012From,Jarke2020Data})} for real-world applicable, secure, and privacy-preserving industrial applications and collaborations have been proposed\new{~\cite{Jarke2020Data,Brauneretal2022A}}. 
One of the main reasons for this situation is that addressing novel use cases is challenging and requires intensive cooperation between industrial companies and cybersecurity experts to come up with suitable use case-fitting solutions.
Today, such cooperation is severely hindered as industrial companies either do not have the required data readily available, are reluctant to share them due to confidentiality concerns, or cannot imagine the possibilities enabled by use case-tailored privacy-preserving building blocks.
Likewise, cybersecurity experts might lack a sufficient understanding of industrial processes, a respective vision of future applications, and the required contacts to significantly advance real-world applications.

To this end, in this paper, we report on our experiences in performing research in the intersection of cybersecurity and industrial application, stemming from several practically applicable research efforts.
Particularly, we present and discuss our complete process cycle from initial bootstrapping to final dissemination as well as challenges and pitfalls that arise when researching at such a practical intersection.
Throughout our discussion, we provide concrete and relatable examples from various published interdisciplinary research in the intersection of cybersecurity and privacy, computer networks and systems, and various industrial domains.

\textbf{Contributions.}
The main contribution of this paper is the presentation and discussion of our methodology for conducting practical research on realizing novel forms of industrial applications and collaboration through recent advances in cybersecurity and privacy research:
We propose a process cycle for cybersecurity research in industrial scenarios, consisting of the identification of a use case, the acquisition of data and its analysis, a research and development phase, and the eventual evaluation and dissemination of results.
Following the complete process cycle of performing such research, we provide an overview of common challenges and pitfalls as a way to bootstrap further research in this emerging field.
\new{%
We further outline the interplay between our process cycle and the formalized data lifecycle in research data management (RDM), underlining their compatibility.}
We illustrate our discussion by using interdisciplinary, real-world examples and experiences involving a variety of stakeholders in the industry.

\textbf{Organization.}
The remainder of this paper is structured as follows.
In Section~\ref{sec:background_relatedwork}, we give information on our interdisciplinary research agenda on creating an Internet of Production~\cite{Pennekampetal2019Towards}, briefly summarize our recent work on real-world use cases from different industrial domains, and revisit today's established scientific methodology.
Section~\ref{sec:design} constitutes the central part of this paper, where we present and discuss in-depth our process cycle view on utilizing cybersecurity and privacy research to realize novel forms of industrial applications and collaboration.
Subsequently, in Section~\ref{sec:parameterexchange}, we further illustrate our process cycle using our recent paper on privacy-preserving parameter exchange within the domains of injection molding and machine tools as a case study~\cite{Pennekampetal2020Privacy-Preserving}.
We discuss further considerations on security research for industrial scenarios in Section~\ref{sec:discussion}, before concluding in Section~\ref{sec:conclusion}.

\section{Our Background \& Related Work}
\label{sec:background_relatedwork}

To go into more detail on the envisioned improvements that we want to realize at the intersection of industrial applications and secure digitization, we introduce our research agenda as a means to identify real-world use cases and establish interdisciplinary collaboration in Section~\ref{subsec:iop}.
In Section~\ref{subsec:papers}, we briefly introduce our recent work covering different real-world use cases, before summarizing from related work why today's established practices and methodologies regarding cybersecurity do not meet all requirements for cross-domain and inter-company research in Section~\ref{subsec:artifacts}.

\subsection{Towards an Internet of Production}
\label{subsec:iop}

As a catalyst for our research efforts in the intersection of cybersecurity and industrial application, we can rely on an established network as a reliable way to identify real-world use cases and bootstrap interdisciplinary collaboration:
Our university's cluster of excellence \emph{Internet of Production (IoP)}~\cite{Pennekampetal2019Towards,Brauneretal2022A} brings together more than \num{200} researchers from more than \num{35} institutes, ranging from various engineering disciplines over computer science to business and social sciences.
We, researchers, cooperate to jointly improve production technology, \eg by implementing cross-domain and inter-company collaborations utilizing data from all parts of a product's lifecycle (\ie production, development, and usage)~\cite{Pennekampetal2019Towards}.
Furthermore, the cluster accelerates our research efforts by providing use cases and production data from more than \num{20} spin-offs and can implement our findings in the production processes of these spin-offs directly using the affiliated entrepreneurship center~\cite{RWTH-Innovation-GmbH2000Offerings}.

While recent advances concerning CPS and the IIoT contribute to promising developments with an application in production~\cite{Xietal2020Virtual,Schlegeletal2020Methodological,Mannetal2020Study,Mannetal2020Connected,Lippetal2020When,Grochowskietal2019Applying,Glebkeetal2019A,Buckhorstetal2021Holarchy,Bodenbenneretal2020Domain-Specific,Beheryetal2020Action,Ayetal2019Kernel,Niemietzetal2020Stamping,Glebkeetal2019Towards,Bibowetal2020Model-Driven,Liebenbergetal2020Information,Brillowskietal2021Know-How,Lippetal2020Flexible,Buschmannetal2021Data-driven,Crameretal2021Towards,Lippetal2021LISSU:,Samsonovetal2021Manufacturing,Xietal2021Tool,Beckeretal2021A,Bodenbenneretal2021FAIR,Brecheretal2021Gaining,Brockhoffetal2021Process,Elseretal2021Potentials,Mertensetal2021Human,Schuhetal2021Development,Kirchhofetal2022MontiThings:,Kunzeetal2021Detecting}, the need for security to enable novel forms of industrial collaboration must be especially considered for real-world deployments, where CPS are likely connected to the Internet.
Apart from operational reasons (\eg to ensure safety or privacy), secure approaches also convince traditionally conservative stakeholders, such as companies that fear for their competitiveness, to contribute to cross-domain and inter-company collaborations.

Given that any improvements proposed within this context depend on collaboration between multiple stakeholders~\cite{Pennekampetal2019Security,Pennekampetal2021The,Pennekampetal2021Confidential}, the resulting dataflows, as well as their security and privacy, are of utmost importance for its success~\cite{Pennekampetal2019Dataflow,Jarke2020Data,Braunetal2020An,Mangeletal2021Data}.
Only by combining security research with any novel form of industrial application, sustainable development can be achieved~\cite{Pennekampetal2021Unlocking}:
While insecure approaches significantly hinder and prevent adoption, novel security building blocks initially often lack real-world deployability, \eg as their usage is not accessible to average users.
Consequentially, connecting both areas through interdisciplinary research is an important step to gradually turn the vision of the IoP into reality.

\subsection{Our Recent Work on Real-World Use Cases}
\label{subsec:papers}

So far, we have tackled the challenge of realizing an IoP from different perspectives and by considering various real-world use cases with diverse needs.
In the following, we briefly present our recent work (labeled with the respective publication venue, \eg IMC for our paper at the Internet Measurement Conference 2020) that serves as the foundation to derive our interdisciplinary research methodology (\emph{process cycle}) that we introduce in this paper.

\textbf{\setpaper{Internet Measurement Conference}{IMC}{paper:imc}.}
Following the idea of getting an overview of the state-of-the-art in ``securely'' deployed industrial applications~\cite{Roepertetal2020Assessing}, we decided to conduct Internet-wide industrial security measurements~\cite{Dahlmannsetal2020Easing}.
For responsible incident reporting, we had to contact and interact with manufacturers and owners of machines that are operated insecurely alike.
Subsequently, we also extended our scans to cover additional protocols~\cite{Dahlmannsetal2022Missed}.

\textbf{\setpaper{Information Processing \& Management}{IP\&M}{paper:ipm}.}
We group our research regarding secure industrial collaborations according to two dimensions:
\emph{along} and \emph{across} supply chains.
Initially, we looked into data sharing \emph{along} supply chains as we could (partially) benefit from well-established trust relationships.
Apart from challenges concerning data reliability and authenticity~\cite{Pennekampetal2020Secure}, we identified the need for improved privacy, especially for dynamic supply chains, as expected in an IoP~\cite{Pennekampetal2020Private}.
We evaluated our proposed architecture with a real-world manufacturer of an electric vehicle~\cite{Baderetal2021Blockchain-Based}.

\textbf{\setpaper{Encrypted Computing \& Applied Homomorphic Cryptography}{WAHC}{paper:wahc}.}
Companies frequently compare their performance with other stakeholders to identify the potential of extended collaborations \emph{along} and \emph{across} supply chains.
Given that such comparisons can leak valuable private data, for example, on the business performance, we revisited existing solutions to (securely) implement company benchmarking~\cite{Pennekampetal2020Revisiting} and proposed a new architecture.
Here, we based our research on a real-world benchmark in the domain of injection molding.

\textbf{\setpaper{Symposium for Database Systems for Business, Technology, and Web}{BTW}{paper:btw}.}
Regardless of data sharing \emph{along} or \emph{across} supply chains, data interoperability between multiple stakeholders (and potentially across domains) is desirable.
Motivated by a use case in the domain of textile engineering, we realized FactStack~\cite{Gleimetal2021FactStack:}, an interoperable data sharing stack, based on the conceptual FactDAG model~\cite{Gleimetal2020FactDAG:}, timestamped URLs as persistent identifiers~\cite{Gleimetal2020Timestamped}, and provenance data~\cite{Gleimetal2020Expressing}.

\textbf{\setpaper{Annual Computer Security Applications Conference}{ACSAC}{paper:acsac}.}
We further proposed a parameter exchange~\cite{Pennekampetal2020Privacy-Preserving} to allow for secure (and private) data sharing \emph{across} supply chains, even with conservative stakeholders.
In this case, we first tackled a use case in the domain of injection molding.
However, we also evaluated a use case covering machine tools to demonstrate that our novel approach is indeed universal.

\textbf{\setpaper{International Conference on Industrial Cyber-Physical Systems}{ICPS}{paper:icps}.}
Before process parameters can be shared, their impact on the process first has to be measured efficiently.
To cope with increasing process data volumes, we leverage in-network computing to realize (simple) computations directly in the network and on dataflows.
In one specific use case in the area of metrology systems, we perform coordinate transformations to map information from locally used coordinate systems at one machine to systems used at other machines~\cite{Kunzeetal2021Investigating}.

Overall, we base the methodology that we are presenting in this paper on our recent work.
In particular, we refer to our past experiences when presenting the individual steps of our process cycle in Section~\ref{subsec:designdetails}.
Additionally, we elaborate on our \ref{paper:acsac} paper as part of our case study (Section~\ref{sec:parameterexchange}).

\subsection{Revisiting Artifacts and Methodologies}
\label{subsec:artifacts}

One of our first steps to solve problems in the industrial domain with mechanisms from cybersecurity and privacy is to look for promising existing solutions and data in the security and privacy domain that might match the requirements of the given use case.
While research methodologies, including artifact badging, support us in getting access to data used and algorithms proposed in research, in parts, these approaches are still insufficient to truly understand the technical background and impact of corresponding work.
Its usability is further challenged by the artifacts' quality as the provided material frequently contains errors or only has a narrow focus.

The trend of artifact badging~\cite{ACM2020Artifact} helps us immensely as it motivates researchers to share data and source code artifacts along with their papers.
Zheng et al.~\cite{Zhengetal2018Cybersecurity} further show that papers with publicly shared datasets tend to receive a broader reception by the research community.
Additionally, mandatory artifact evaluation creates larger collections of artifacts that are easily accessible and can be reused~\cite{Collbergetal2015A}.
However, the desirable goal of facilitating access to the actual knowledge hidden inside the publications and artifacts, \ie interconnecting the knowledge, still requires research~\cite{SEARCCH2020Sharing}.

Furthermore, badging alone is insufficient to fully understand the capabilities and limitations of an existing solution.
For example, Zilberman~\cite{Zilberman2020An} finds that an artifact warranting the ``reusability'' badge does not necessarily suffice to convey all architectural, implementation, and evaluation limitations of an award-winning ACM SIGCOMM paper.
On a similar note, van der Kouwe et al.~\cite{Kouweetal2018Benchmarking} find that benchmarking ``crimes'' are quite frequent at the top four security conferences, with only a few papers being free of them:
\new{%
A commonly found category is \emph{selective} benchmarking which, for instance, includes choosing evaluation scenarios without proper justification, thereby possibly hiding worst-case scenarios from the readers.
Likewise, Arp et al.~\cite{Arpetal2022Dos} report similar findings, \eg sampling bias in collected data, based on their survey that covers the evaluation methodology of machine learning approaches at top security conferences.}
Above all, primary reasons are the general expectation of significant scientific advances in every publication as well as a diminished value attributed to reporting failures~\cite{Longstaffetal2010Barriers}, although first initiatives also encourage discussing negative results~\cite{PerFail-20222022PerFail}.
Consequently, even if artifacts are evaluated, badged, and available, they might still not behave as reported, and the entire scope of their functionality possibly cannot be easily grasped.

While the badging process itself might be adjusted with additional evaluation criteria accounting for the weaknesses of the presented solutions~\cite{Zilbermanetal2020Thoughts}, there is a commonly identified need for research guidelines making it easier for researchers to avoid common mistakes.
One well-known example is the evaluation checklist of ACM SIGPLAN~\cite{Bergeretal2018SIGPLAN}, but other work also focuses on directly strengthening the underlying methodology~\cite{Evaluate-Collaboratory2010Experimental,Balensonetal2015Cybersecurity}.
\new{%
Significant efforts in this context also tackle conducting and evaluating user studies in the security realm~\cite{Garneauetal2016Results,Kroletal2016Towards,Dykstraetal2018Cyber,Ben-Salemetal2011On,Levesqueetal2014Computer}.
Another branch of work focuses on sound (and realistic) testbed-based experiment strategies, which become increasingly important with more and more complex systems, \eg in the context of CPS~\cite{Linetal2020Cyber-Physical,Mathuretal2016SWaT:,Carrolletal2012Realizing,Crusselletal2019Lessons,Dumitrasetal2011Experimental,Greenetal2017Pains}.}

\new{%
Despite these steps, true interdisciplinary work, such as our collaborations in the IoP, posits novel challenges that are not yet addressed by existing guidelines.
In particular, substantial knowledge gaps between collaborators of different domains are frequently present, which stands in contrast to initiatives, \eg by the NSA~\cite{Dykstraetal2019Lessons}, that push information sharing among knowledgeable peers.
Consequently,} transitioning the solution space of the cybersecurity and privacy domain to practical problems, such as in the industrial domain, needs to be facilitated~\cite{Maughanetal2013Crossing}.
In an effort to ease collaborations for future work, we thus present and discuss our methodology for conducting interdisciplinary research.

\section{A Process Cycle on Collaboration}
\label{sec:design}

Based on our experiences when realizing novel forms of industrial collaborations at the intersection of security research and real-world use, we derived a process cycle that allows researchers to organize their work accordingly.
First, in Section~\ref{subsec:overview}, we provide a high-level overview before detailing the individual steps in Section~\ref{subsec:designdetails}.
Finally, in Section~\ref{subsec:takeways}, we look at additional lessons learned to support future research at this challenging intersection.

\begin{figure*}[t]
  \center
  \includegraphics{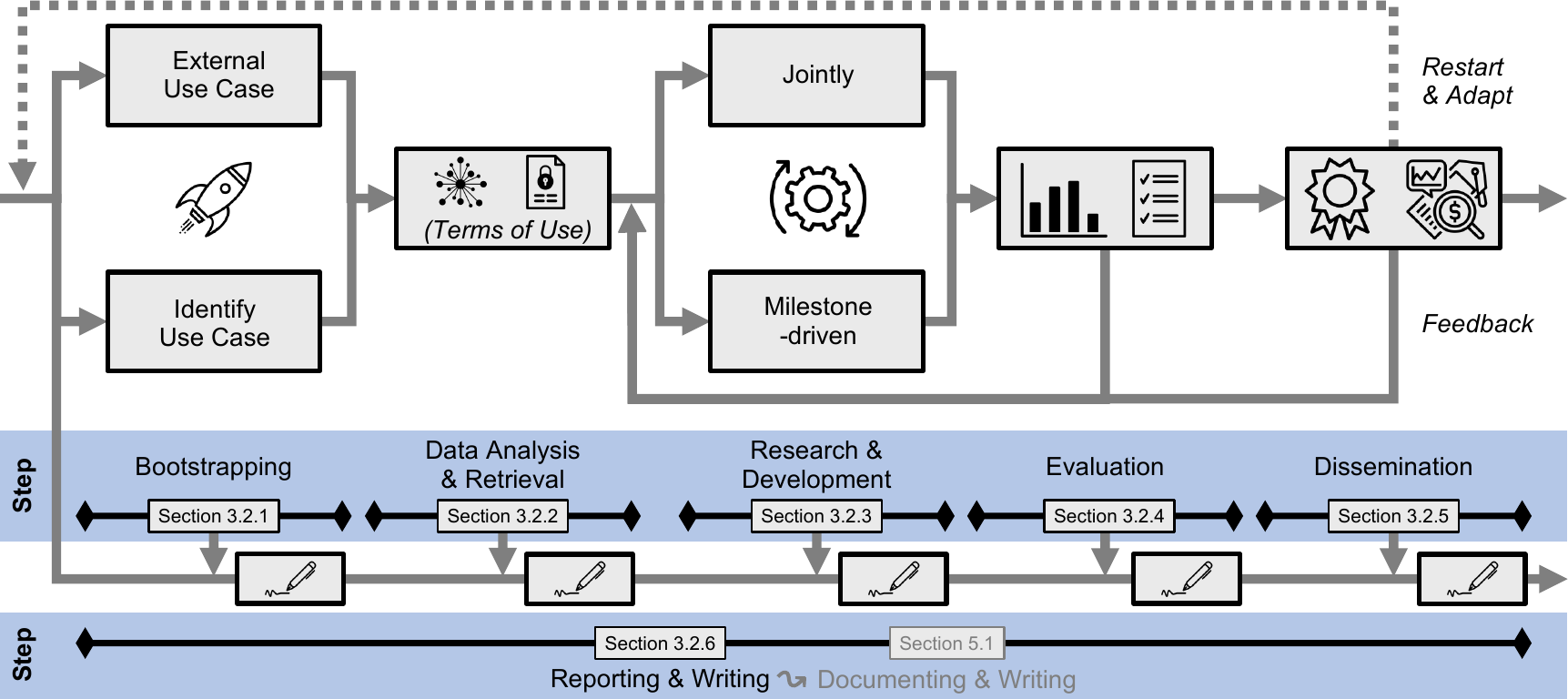}
  \caption{%
    We derived an abstract process cycle consisting of six individual process steps to formalize the different phases of an interdisciplinary research project at the intersection of cybersecurity research and industrial applications.
    While the majority of steps, \ie \hyperref[subsubsec:bootstrapping]{\emph{Bootstrapping}}, \hyperref[subsubsec:data]{\emph{Data Analysis \& Retrieval}}, \hyperref[subsubsec:research]{\emph{Research \& Development}}, \hyperref[subsubsec:evaluation]{\emph{Evaluation}}, and \hyperref[subsubsec:dissemination]{\emph{Dissemination}}, build upon each other, the \hyperref[subsubsec:writing]{\emph{Reporting \& Writing}} step is an accompanying phase.
    In Section~\ref{subsec:rdm}, we further describe how to enhance ($\bm{\leadsto}$) the accompanying step to a \hyperref[subsec:rdm]{\emph{Documenting \& Writing}} phase by incorporating practices from research data management (RDM).
  }
  \label{fig:processcycle}
\end{figure*}

\subsection{Design Overview}
\label{subsec:overview}

Based on our previous work (cf.\ Section~\ref{subsec:papers}), we were able to identify patterns that repeatedly occurred as part of our research progress.
We combined these experiences and derived a (universal) process cycle that is applicable to research striving for secure and private real-world applications in the IIoT.
As such, this process cycle nicely fits to work that advances industrial applications, such as our agenda of realizing an Internet of Production (cf.\ Section~\ref{subsec:iop}).

Overall, we separate the cycle into \emph{six} distinct steps, as we illustrate in Figure~\ref{fig:processcycle}.
The individual steps are as follows.

\textbf{Bootstrapping.}
First, a suitable use case to work on must be selected.
Here, an externally requested use case can be chosen, or alternatively, a use case can be identified independently.
In the latter case, a significant challenge is to make sure that the selected use case is a relevant real-world application, \ie it is worth pursuing, as well as evaluable and deployable in realistic industry settings.

\textbf{Data Analysis \& Retrieval.}
Second, researchers must make themselves familiar with any available data (sources).
This process can also cover the reuse of data.
Having access to meaningful information is key to enabling solutions tailored to the use case.
To this end, researchers might even have to accept specific terms of use or sign a non-disclosure agreement to comply with data usage policies that are in place to protect sensitive (real-world) use case data.
Additionally, apart from understanding the semantics, data usually must be parsed and pre-processed for the subsequent development.

\textbf{Research \& Development.}
Third, a suitable approach for the current use case must be developed.
Depending on the type of collaboration, researchers and the use case provider can work jointly, \ie they have a close feedback loop.
Alternatively, the process can be milestone-driven, where progress is only reported periodically.

\textbf{Evaluation.}
Fourth, researchers must look at the applicability and feasibility of their proposed approaches.
Using the (real-world) data, they can conduct realistic, real-world-motivated evaluations \new{that convey practical implications}.

\textbf{Dissemination.}
Fifth, the dissemination \new{of research results, conclusions, lessons learned, methodologies, artifacts, and datasets} is an essential step to raise the awareness of this area, to encourage additional research, and, most importantly, to further utilize the developed approach.
In this context, established practices (cf.\ Section~\ref{subsec:artifacts}), such as public artifacts sharing, come to mind.
Especially w.r.t.\ the \emph{data analysis \& retrieval} process step, this phase has a significant impact as publicly-shared (and ideally independently audited) artifacts can be reused in other contexts and use cases.
Apart from being a well-acclaimed contribution, sharing artifacts also supports incremental research as the tiresome task of repeating evaluations or reimplementing previous concepts is avoided, while the overall comparability of results is improved.
Thus, it eases the use of standardized methodologies.

\textbf{Reporting \& Writing.}
Finally, we identified a step that accompanies all previously presented process cycle phases.
On the one hand, researchers have to report on their recent steps to drive the discourse with their collaborators.
On the other hand, they have to keep the (paper) writing in mind as one of the main ways to receive external feedback (\eg through peer-review) and to push the dissemination of their newly developed approaches.
Most notably, in this step, both researchers and collaborators are involved alike.

Once this process cycle has been completed, another use case can be tackled while also incorporating the newly generated knowledge and experiences.
As such, every past use case also influences future challenges and potentially contributes to their resolutions.

Next, we detail the individual steps of our process cycle.

\subsection{Revisiting the Process Cycle}
\label{subsec:designdetails}

Following this abstract overview on our process cycle, we now elaborate on the individual steps as well as their underlying aspects and refer to our experiences that we collected as part of our research (cf.\ Section~\ref{subsec:papers}).
Thereby, we \stephighlight{highlight} major alternatives of process steps, which we also include in Figure~\ref{fig:processcycle}.
Subsequently, as part of our case study, which we present in Section~\ref{sec:parameterexchange}, we explicitly discuss all process steps of our recent \ref{paper:acsac} paper in more detail.

\subsubsection{Bootstrapping}
\label{subsubsec:bootstrapping}
The process of tackling a relevant real-world problem must start with finding such a problem before envisioning a solution for improvements.
Although both steps usually require specific industrial domain knowledge, researchers can choose to actively collaborate with external practitioners (\emph{External Use Case}) or bootstrap the work on their own (\emph{Identify Use Case}).
Here, the main challenges are that researchers and practitioners might not share the same visions, and they might not immediately understand each other on a technical level.
Thus, both the language and the ideas of the involved research directions must be carefully bridged.

\stephighlight{External Use Case}
As part of our past work, we were in talks with practitioners from the domain of injection molding to look at their data security-related research problems.
Based on this connection, we jointly identified the challenges that we addressed in \ref{paper:acsac} and \ref{paper:wahc}, \ie a lack of sufficient, yet real-world feasible data security for industrial use.
While today's security possibilities were mostly unclear for the practitioners, we as researchers were unable to estimate their needs accurately.
A particular challenge in finding a suitable solution is that conservative companies may lack the vision to advance the state of the art using external data.

\stephighlight{Identify Use Case}
Regardless, research can also succeed without a practitioner that helps to outline a problem.
For example, our work on data sharing improvements along supply chains (\ref{paper:ipm}) is based on our (independent) analysis of related work.
Thus, instead of directly talking with practitioners, we only ingested their views indirectly.
However, as we already experienced during previous research efforts~\cite{Hilleretal2018Secure, Serroretal2020QWIN:}, this step is cumbersome and might result in identified research gaps that do not fit real-world industry needs.
Hence, all available means must be taken to verify the open problem:
Therefore, for \ref{paper:ipm}, we conducted expert interviews on a small scale after having postulated the research gap internally.

Finally, as part of this step, a thorough analysis of related work is needed.
This analysis is not limited to suitable existing solutions, but more importantly, researchers should also try to identify similar (real-world) use cases to improve the universality of their solutions.
Especially when considering today's expectations by academia (cf.\ Section~\ref{subsec:community}), the chances for publication are improved.

\subsubsection{Data Analysis \& Retrieval}
\label{subsubsec:data}
After the problem space has been explored, any needed data must be discovered, shared, and analyzed for later use.
This step also includes signing potentially required usage agreements with companies.

\textbf{Discovery.}
First, researchers have to discover which data is available and then identify which parts of it are relevant.
Afterward, they must check whether this information is also accessible for research use and under which terms.
For example, further use of the benchmarking data (evaluated as part of \ref{paper:wahc}) was limited to use by the practitioner due to its sensitivity.
Again, researchers should be open and try to gather as much information as possible, given that practitioners might not see the relevance of seemingly insignificant data.

Instead of requesting data from practitioners, researchers can ideally rely on public datasets to foster reproducibility by and comparability to other work.
However, depending on public datasets, in contrast to getting access to relevant use case-specific data from collaborators, introduces its own caveats.
For instance, the number of available datasets for cybersecurity research of industrial systems is limited, and the available datasets are usually either not a perfect fit for the use case at hand or only contain poor documentation of the underlying setup~\cite{Contietal2021A}.
The scarcity of datasets partially follows from the fact that the behavior of CPS is best captured by physical testbeds that are expensive to setup and maintain, and partially from the mindset that the generation of a high-quality dataset is not very rewarding for researchers and practitioners alike when weighing the required effort~\cite{McLaughlinetal2016The,Contietal2021A}.
Finally, scaled-down testbeds are also not necessarily representative of larger, real-world systems, which they intend to mimic, further limiting the credibility of these datasets~\cite{Anietal2021Design}.
These general challenges of today's cybersecurity research on CPS can be illustrated using the example of industrial intrusion detection systems:
Many novel ideas are evaluated using datasets of a single entity, \ie the iTrust research center at the Singapore University of Technology~\cite{Singapore-University-of-Technology-and-Design2020iTrust}, resulting in a natural bias during all conducted research steps~\cite{Wolsingetal2020Poster:,Wolsingetal2021IPAL:}.

Thus, as incorporating the ``correct'' data is crucial, we urge researchers to carefully review their selection process and to expect conceptual flaws due to the origins of (real-world) data.

\textbf{Sharing.}
Following the discovery, when data is not publicly available, the information must be transferred to the researchers.
Most notably, the main issue is usually not whether data exists, but whether it could be shared for external research purposes, \ie privacy concerns challenge its use by researchers.
Thus, any questions concerning the terms of use must be (formally) resolved (see below regarding their implications on our cycle).

Our work in the area of industrial collaboration is especially challenged by the fact that receiving real-world data from different companies is mostly impossible as they fear data leaks and a loss of their competitive advantage.
Thus, we usually have to split existing data or ``reuse'' the same dataset for all parties in our setting.

A related aspect that is especially prevalent in industrial settings is the challenge of real-world implications, \ie does our data use have any implications on productive systems.
For example, for our Internet measurements (\ref{paper:imc}), we directly interacted with real-world industrial deployments.
Thus, following best practices and ethical considerations, we had to ensure that our work had no real-world implications, mainly concerning safety and security.

\textbf{Analysis.}
After having access to the data, understanding its semantics and structure is the next challenge.
Apart from identifying gaps in the available information, the use case data must be translated into a proper form, where unnecessary information is removed.
Given that practitioners might not be familiar with the used data sources either, this process can become quite time-consuming until a correct pre-processing has been applied.
In addition to accurate insights, this step also allows researchers to transform the information into representations in open and standardized formats.

Naturally, a correct understanding of the available information is key to avoid subsequent errors.
However, achieving this state is difficult as often no or little documentation is available.
Industry data frequently originates from proprietary systems that impose additional obstacles to this step.
Already the first glance might be misleading, which invalidates all subsequent steps of the research.

\stephighlight{Terms of Use}
As stated before, industrial companies are notoriously conservative due to their competitive standing:
Data is both valuable (also for global improvements) and sensitive at the same time.
Therefore, we noticed that discussions regarding potential improvements fueled by security research are usually very enthusiastic.
However, when it comes to sharing any data, they are frequently reserved.
Thus, researchers might be required to accept any terms of use and/or to sign a non-disclosure agreement.
Different best practices in companies and academia can further delay this already slow process.

Researchers should explicitly be aware of any implications for the publication process, such as mandatory approval processes, constraints in reporting (negative) findings, or licensing aspects concerning the open-sourcing of (research) artifacts.

For our \ref{paper:ipm} approach, we did not have to rely on their data for the development as we already generated artificial datasets to model real-world supply chains for our previous evaluation.
Thus, we did not encounter any delays from the obstacles of signing a non-disclosure agreement.
However, for other settings, coming up with realistic or even usable artificial datasets might not be an option.
Thus, getting access to otherwise confidential (protected) use case data unlocks otherwise unavailable research challenges.

Another angle on this aspect is the use of experimental hardware for prototype development.
For example, as part of our work on in-network coordinate transformations (\ref{paper:icps}), we deployed an Intel Tofino switch~\cite{Intel2020IntelR}.
Currently, the use of Tofino is, among other things, still tied to a mandatory approval process for publications with Intel engineers.
While this step complicated the submission process due to an increased number of involved parties, Intel's helpful feedback allowed us to further improve our paper.

\subsubsection{Research \& Development}
\label{subsubsec:research}
Once the needed use case data is prepared, work on the solution can start.
From a research perspective, this step does not differ from usual research.
However, we observed that the way of interaction with the use case partner can affect the progress.
In particular, we distinguish a joint development approach from a more milestone-driven paradigm.

\stephighlight{Jointly}
Collaboratively conducting an agile process is extremely helpful to correct any newly occurring (or remaining) misconceptions early on.
Furthermore, this approach allows researchers to demonstrate any incremental progress while also raising awareness of the associated technical challenges.
Thus, it fosters the discourse.

Despite a thorough data analysis \& retrieval phase, we still discovered a misconception concerning the use case data of our parameter exchange (\ref{paper:acsac}).
Fortunately, our close collaboration quickly allowed us to adjust the development accordingly (cf.\ Section~\ref{sec:parameterexchange}).

\stephighlight{Milestone-driven}
Alternatively, reporting on progress can be limited to specific (previously communicated) milestones.
Here, the benefit is that the practitioner can focus on the use case and is not repeatedly distracted by underlying technical details.
However, receiving timely and accurate feedback on the recent progress is more challenging when following this organizational paradigm.

For our supply chain architecture (\ref{paper:ipm}), we finished an initial prototype without having access to any (realistic) use case data.
Thus, we had no feedback regarding the real-world applicability (or even correctness) of our approach.
Apart from the risk of solving the wrong real-world ``problem'', we noticed that obtaining suitable evaluation data can become extremely tiresome, for example, if a non-disclosure agreement needs to be signed.
Besides, a lack of real-world data, \ie artificial examples only, especially complicates the subsequent publication of the work (cf.\ Section~\ref{subsubsec:writing} and~\ref{subsec:community}).

Overall, we want to highlight that having a feedback loop in place is very beneficial, given that any project at this intersection puts a strong emphasis on real-world applicability.
Hence, ensuring that this key goal is met also helps to confirm the correctness of the pursued solution without any additional overhead.
Finally, the practitioners will feel more integrated into the research project, which reduces the risks of indifference or dissatisfaction.
In the past, we heard reports from practitioners in similar interdisciplinary projects that they feel to just serve as data sources, \ie they felt underrated concerning their contribution to the research progress.

Another major challenge that we repeatedly came across as part of our work concerns the scalability requirements of the approach in question.
While the evaluation itself follows as the next step in our process cycle, realistic constraints are essential upfront to come up with a fitting solution.
We frequently noticed that the exact future needs are still unclear as the overall future development concerning data sharing benefits and data security demands at this comparably novel research intersection are mostly uncertain.
Paired with the reservation towards change in these usually conservative environments, correctly inferring the scalability needs is difficult.

\subsubsection{Evaluation}
\label{subsubsec:evaluation}
In addition to traditional evaluations that are part of every research project, we explicitly want to highlight the need to check for real-world applicability.
Overall, every evaluation should indicate whether the developed prototype is suitable to tackle the targeted real-world problem and which consequences the results entail.
If needed, the developed approach must be revised thoroughly according to new findings during the evaluation.

As the main goal of the evaluation is to show that the developed prototype meets real-world requirements and to avoid any inaccurate conclusions, we recommend relying on real-world use case data at all times.
While already traditional evaluation in the privacy research area can take significant time, high volumes of real-world data can increase the time beyond that.
Rather, however, the need to access or interact with industrial machines (that are used in production) may impact the duration of this phase.
Naturally, as in all other steps, specific caution must be exercised concerning possible safety aspects and environmental impact~\cite{Henze2020The}.

As we elaborate in more detail in Section~\ref{subsec:community}, evaluating all proposed approaches with a convincing real-world use case is highly beneficial for the publication process.
To further improve the impact of publications in terms of security research, we recommend conducting use case-independent evaluations as well, \ie to generalize the security contributions as much as possible.
This aspect supports researchers who are challenged with deriving claims that are universally valid or collecting all-encompassing empirical evidence.

\subsubsection{Dissemination}
\label{subsubsec:dissemination}
Especially with the focus on real-world applicable solutions, the step to disseminate the progress is crucial.
Here, we identified different aspects where the interests of researchers and practitioners often diverge, \eg the usability of a prototype developed during research or publishing use case data.
Hence, researchers must discuss and agree on all these aspects with all involved stakeholders early on in the process cycle to avoid any misunderstandings.
Most notably, open-sourcing both software and use case data must be discussed right at the beginning as it can also influence the initially agreed-upon terms of use (cf.\ Section~\ref{subsubsec:data}).

\textbf{Readiness Level.}
Keeping in mind that research is usually only interested in developing proof-of-concept prototypes, the trade-off between their usability and the impact for research needs specific consideration.
Especially with practitioners as partners who strive for real-world deployments of said developments, the expectations should be clarified at the first start of each process cycle.
We believe that the contribution of opening practitioners for novel approaches that are enabled by security and privacy research is already of intangible value despite a potentially limited product maturity.

\textbf{Preparing Artifacts.}
Orthogonally, the publication of datasets is a delicate aspect as they can still contain sensitive use case information.
As removing all critical, potentially insights-leaking features is very challenging, we can understand the reservations of (some) companies when it comes to their otherwise private data.

Even when publishing consolidated datasets of already publicly available data, for example, as gathered during Internet-wide measurements (\ref{paper:imc}), researchers are responsible for ensuring that the data does not contain privacy-sensitive user data or any information on identities, and cannot be misused as information source for attackers.
To this end, for the publication of our \ref{paper:imc} dataset, amongst other measures, we mapped IP addresses to a numeric space, replaced hostnames, and removed received payload data.

\textbf{Reusability.}
In addition to verifiability, the goal of published artifacts should be to improve reusability.
However, preparing an all-encompassing README is far from trivial.
This process is challenged even more if arbitrary (external) use case data should be supported as the respective data sources can vary significantly in terms of both syntax and semantic.
Here, domain-induced misconceptions might challenge any attempts to pre-process data correctly.

The research community already looks into ways to improve this situation and offers programs, such as artifact evaluations, badges, and other awards (cf.\ Section~\ref{subsec:artifacts}).
For example, for our parameter exchange (\ref{paper:acsac}), we open-sourced the fully-tested implementation and all use case data~\cite{COMSYS2020Privacy-Preserving} and received a functional badge~\cite{Pennekampetal2020Privacy-Preserving}.
With a large number of available artifacts, the likelihood of defining (and reusing) a standardized research methodology across use cases, domains, and academia increases.
As for all research in general, advances building upon existing approaches can, in the long run, also help to tackle problems that seem unsolvable at the moment.

\textbf{Responsible Disclosure.}
Nevertheless, when researchers discover alarming information in their data, ethical principles require not only to publish the data, but also to actively and responsibly disclose the relevant findings.
For our Internet-wide analysis of the security configuration of OPC Unified Architecture (OPC UA) servers (\ref{paper:imc}), we reached out to all relevant operators if possible.
We informed the operators of systems with problematic security configurations whenever we were able to identify them via email (and, in one case, via phone).

\textbf{Bootstrapping Further Research.}
All these different aspects of the dissemination can help to ensure progress.
Eventually, we are confident that any work in this novel research intersection can encourage additional work, resulting in a larger overall acceptance of this challenging, yet practical area.
Regardless, with any finished project, researchers can now revisit other use cases and build upon the newly collected experiences.
Thus, the ability to also work on incremental research is improved.
Due to the goal of real-world applicable approaches, reporting on negative findings, as encouraged by the LASER workshop series~\cite{LASER2011The}, is very supportive for the cybersecurity research community.
For example, the idea of our OPC UA Internet measurements (\ref{paper:imc}) was sparked by our work to offer a tool to assess the security of OPC UA deployments~\cite{Roepertetal2020Assessing}.

\subsubsection{Reporting \& Writing}
\label{subsubsec:writing}
In addition to the applied research in terms of analyzing data and programming software, reporting on the research progress and publishing papers is an important part of research.
We consider this process step to evolve in parallel to all previously presented steps as each step provides meaningful input. 

Further, we noticed various challenges with scientific writing when many stakeholders from different domains are involved.
Each stakeholder is used to their individual best practices when it comes to reporting and writing.
For the paper writing, the most prominent difference we noticed is the use of traditional office client applications in contrast to the use of \LaTeX{} in research institutes.
Likewise, version control systems, such as Git, were not commonly used.
Consequently, an adjustment by a subset of involved people is usually needed.
However, the challenges also comprise additional expectations regarding the (writing) style, the submission process, and other organizational matters.
Thus, this process step must be approached with caution to address the expectations of everyone involved.
We will look into the challenges of finding a suitable venue and research community in more detail in Section~\ref{subsec:community}.

An aspect that we want to raise here explicitly is the approval process that might be completely different from traditional work, \ie all parts of the work or paper have to be approved.
Therefore, a separation of responsibilities is usually not accepted.
This aspect is also relevant as part of negotiated terms of use or non-disclosure agreements (cf.\ Section~\ref{subsubsec:data}) as they might introduce specific rules for all parties.
Ultimately, these constraints challenge the ability to work in parallel and complicate last-minute changes.
With the required approval for all changes, endless feedback loops might be the result, which is especially problematic for paper submissions that have a fixed deadline.
Thus, all expectations and deadlines should be communicated clearly and early on.
For our \ref{paper:acsac} paper, we eventually dealt with a total of nine authors from three departments, all contributing their own publication cultures and expectations.

In Section~\ref{subsec:rdm}, we will elaborate on our vision to form this step into a \emph{Documenting \& Writing} phase by looking into the potential impact of an applied research data management (RDM).
In our previous interdisciplinary efforts, we did not follow the associated RDM practices in detail, and as such, we cannot report on their impact or talk about their impact on our derived process cycle.

\subsection{Takeaways \& Other Lessons Learned}
\label{subsec:takeways}

In addition to all the aspects that are attributable to specific steps in our process cycle, we also experienced some additional lessons learned that do not fit into a single step.
In the following, we discuss these aspects and try to postulate corresponding takeaways.

\textbf{Communication.}
As our research focuses on interdisciplinary research topics, an active exchange between all involved stakeholders is very important.
While the first discussions are challenging to master, the situation will improve over time as the awareness of the motivation, challenges, and fears is increasingly understood by the collaborators.
The goal must be to tackle the issue jointly rather than working individually and merging the results in the end.
Thus, we want to stress that communication is important to resolve implicit assumptions that potentially hinder substantial progress.

\textbf{Curiosity.}
Even though a domain expert is usually involved as part of the collaboration, we noticed that challenging their views, assumptions, and ideas is helpful to deepen the understanding of the topic on the one hand and to revisit the fit of a chosen approach on the other hand.
Hence, we recommend questioning everything and not taking anything for granted:
Bridging the domains is challenging and takes time, but especially, the different (uncommon) views on today's problems or situations help to develop suitable practical solutions or even to identify novel applications.

\textbf{Artifacts Reuse.}
Generally, artifacts reuse is an important aspect (and issue) in science (cf.\ Section~\ref{subsec:artifacts}).
However, to judge the real-world applicability of a new approach, respective data is imperative.
Unfortunately, due to privacy concerns and its sensitivity, real-world industrial use case data is a scarce good and thus hinders not only the evaluation of improved solutions, but also challenges their development in the first place.
Even if artifacts are available, the supplied documentation is only minimalistic, which challenges the applicability in related use cases.
Overall, we identify the need to rethink the way artifacts and datasets are shared for research, especially to better support the development of real-world applicable approaches.
For example, \new{as implemented in SEARCCH~\cite{SEARCCH2020Sharing},} a global reputation system\new{, supporting elaborate reviews,} that explicitly focuses on (and rates) the quality of shared datasets \new{and research artifacts} could be used.

\textbf{Results Comparison(s).}
As indicated before, evaluating approaches with real-world data is a significant challenge for such applied research.
However, concerning the scientific discourse, comparing newly proposed solutions to other previous work is also very difficult as either no directly fitting approaches are out there, as no datasets are publicly available, or the source code artifacts are not publicly shared.
The latter issue also challenges incremental research of seemingly addressed use cases by other researchers.
Unfortunately, any artifacts sharing is greatly hindered by the competitive mindset of involved stakeholders.
In the long run, the academic research community has to rigorously change their mindset accordingly, so that publishing all relevant material and use case data is widely-accepted and the default practice.

Overall, we have identified aspects that are crucial for individual collaborations (communication and curiosity) and highlighted issues that require significant changes for work in this interdisciplinary research area (artifact reuse and results comparisons).
Regardless, our past work underlines that impactful and real-world applicable research for industrial use can already be conducted successfully in today's landscape with external collaborators.

\section{Case Study: Illustrating the Cycle Based on a Single Research Project}
\label{sec:parameterexchange}

In the following, we demonstrate the different steps of the process cycle in the form of a case study, using our \ref{paper:acsac} paper as an example.
The paper proposes a parameter exchange to privacy-preservingly enable dataflows of relevant production data between (mutually distrustful) stakeholders in industrial settings.

\textbf{Bootstrapping.}
The work on this specific project was initiated by chance as it followed from a conversation with a practitioner of our cluster of excellence~(cf.\ Section~\ref{subsec:iop}).
During this conversation, the lack of a privacy-preserving exchange mechanism for process parameters came up, which currently hinders the development of process optimization techniques in the domain of injection molding (cf.\ External Use Case).
We noticed conceptual similarities w.r.t.\ privacy concerns between the described use case and our existing work~\cite{Dahlmannsetal2019Privacy-Preserving,Ziegeldorfetal2017BLOOM:} and sought a dedicated meeting to discuss the potential use case and its associated challenges in detail.

The first meeting was dominated by the practitioners giving us an understanding of the idea, its practical relevance, and the key requirements for the envisioned mechanism, \ie we had to gain a sufficient understanding of the target domain to determine suitable approaches.
Based on this information, we filtered our initial ideas on solving this specific challenge.
For example, we rejected one of our ideas as the foundational approach~\cite{Ziegeldorfetal2017BLOOM:} was not applicable for a parameter exchange in the domain of injection molding, mainly because the required matching algorithms in this domain are too complex to be handled efficiently by homomorphic encryption.
Subsequently, we pushed the remaining ideas, one of them seeming especially promising, which was based on previous work~\cite{Dahlmannsetal2019Privacy-Preserving}.
In a second meeting with the practitioners, we presented this specific idea and got consent that it should meet all use case-specific requirements.
Afterward, we continued with the data analysis.

\textbf{Data Analysis \& Retrieval.}
Following the second meeting, the practitioner provided us with genuine data, sourced from the domain of injection molding.
We were allowed to use this data for our prototype and later for the evaluation.
However, the practitioner provided the data only as a text file containing a JSON representation of a Python dict and a pickled file with limited documentation.
Due to communication delays, we were required to reformat the given data, and derive as much information as possible on our own.
Nevertheless, working with the data on this lower level allowed us to formulate more specific questions later on.
Later, we obtained additional documentation that illustrated the origin of the provided data and the relevance of most technical terms, \ie it allowed us to comprehend the use case even as a researcher from another domain.
This progress turned out to be crucial for the success of the project.

Considering the data only as a set of values and using them obliviously is precarious as it may lead to far-reaching misunderstandings that might yield problems with key components of the solution design.
To exemplify, the provided dataset contained three keys for each object in the set: \emph{x}, \emph{y}, and \emph{geometry}.
From the explanations and the correspondence with the practitioner, we initially understood that \emph{x} and \emph{geometry} represent the identifying parameters, while the \emph{y} values represent the output.
However, during the subsequent \emph{Research \& Development} phase, when presenting a first demo of the prototype to the practitioner, following a discussion, we (finally) understood that \emph{x} and \emph{y} represent output values and only the \emph{geometry} is relevant for the parameter identification.
Fortunately, the inevitable modifications to our prototype resulted in few changes only.
However, a closer collaboration with the practitioners, when we initially analyzed the data, could have prevented this misunderstanding.
Here, we noticed that any implicit assumptions can have a critical aftermath concerning the pursued solution.

Subsequently, with the revised interpretation of the data, we designed a sensible input format for our prototype and implemented a pre-processing script that transforms data from the format provided by the practitioner into this input format.
In a later meeting, we presented and reconciled our understanding of the data to the practitioner to ensure that the chosen processing format does not omit crucial information.
Afterward, this phase was concluded.

\textbf{Research \& Development.}
After determining a suitable data format, we implemented a prototype, conceptually following the idea of our previous work.
We agreed on a fixed interval for meetings to present the current state of our work to the practitioner (cf.\ Jointly).
Thereby, we ensured that the practitioner can verify that the functionality meets the requirements.
Moreover, the feedback loop allowed us to address any remaining misunderstandings early on during the development.
For example, we noticed the input data issue, mentioned as part of our description of the previous phase, at the first development meeting.
If this mistake would only have been noticed during the evaluation or the subsequent dissemination, the costs to address it would have been significantly higher.

Finally, these regular meetings led to a good integration of the practitioner and repeatedly opened the room for discussions concerning the overall use case progress, \ie we could easily revisit the initially declared use case requirements.
Following a final presentation of the completed prototype, we began with the evaluation.

\textbf{Evaluation.}
For our first measurements, we utilized the initially provided data from the domain of injection molding.
Unfortunately, we noticed that the approach on which we had focused the most (\emph{PPE}, based on private set intersections) is not scalable to all queries occurring in the real world.
Hence, we decided to implement a second design variant (\emph{BPE}, based on Bloom filters) that is more generally applicable at the cost of slightly reduced privacy.
This example illustrates that the development and evaluation phases cannot be completely separated but are somewhat interlinked, as certain evaluation results might lead to modifications of the design.

For further evaluation, we requested additional, larger datasets.
However, creating the desired data required a time-consuming process with industrial injection molding machines such that we could not obtain more data on time.
To evaluate our implementation with larger datasets despite this limitation, we generated artificial data based on the provided genuine data, utilizing a similar range of values.
We asked the practitioner to confirm its similarity to real-world data to ensure a sound and impactful evaluation.

Meanwhile, we also contacted other practitioners in the domain of mechanical engineering to check whether the targeted use case would be of relevance as well.
Apart from demonstrating the applicability in the domain of machine tools, this procedure allowed us to underline the universality of our design as it is beneficial in different domains as well.
We received a second real-world dataset and had to repeat most previous steps:
Again, we had to understand, pre-process, and double-check the respective use case data.

\textbf{Dissemination.}
Apart from a scientific publication, we planned to publish our fully-tested source code early on.
In particular, our target venue included an artifact evaluation upon acceptance of the publication.
Thus, we eventually had to polish our source code and its documentation.
Maintaining a proper code quality during the implementation, with up-to-date documentation and setup instructions, significantly eased the preparation of the artifacts.
Here, we want to stress that an early decision concerning any artifacts incentivizes proper documentation during development.

To also include the data used for the evaluation in the publication, the coordination with the practitioners, who contributed the data, is crucial so that all private information from the datasets can be removed before publication.
For instance, our dataset contained company names that are used colloquially as synonyms for products such that we had to remove these references prior to open-sourcing.

\textbf{Reporting \& Writing}
We decided on \ref{paper:acsac} as our target venue early during development to clarify the requirements, deadlines, and the available timeframe.
\new{%
Given its explicit focus on applied security, \ref{paper:acsac} constitutes a prime candidate when looking for a venue that is interested in our applied research with its real-world use cases.}
For a detailed description of the considered use cases, we asked the practitioners who provided the datasets to participate during the writing.
The collaboration with researchers from other domains and departments introduces certain challenges.
First, the unfamiliarity with \LaTeX{} required us to, at least partially, work with Microsoft Word.
Second, strict practitioner-driven publication requirements required us to finish the paper significantly earlier than demanded by the venue to allow for hierarchical approval workflows.
To work around these challenges, communication of different publication cultures and expectations is beneficial as stakeholders might be unaware of other ways.
Eventually, and regardless of any interdisciplinary-induced challenges along the complete process cycle, we successfully published our paper at \ref{paper:acsac} with nine authors from three departments.
We further contribute an evaluated artifact of source code and real-world datasets.

With this case study in mind, in the remainder of this paper, we now focus on challenges that arise from the different viewpoints of industry and academia, especially concerning real-world deployments and (scientifically) disseminating the obtained results.

\section{Remaining Process Considerations}
\label{sec:discussion}

Building upon our derived process cycle, we now focus on related challenges that significantly impact on our used methodology.

First, in Section~\ref{subsec:rdm}, we highlight the benefits that the inclusion of research data management (RDM) could have for the process cycle and, in turn, for our work.
So far, we have neglected these respective practices despite their potential for (research) artifacts.

Subsequently, we focus on the views of industry and academia in Section~\ref{subsec:deployments} and~\ref{subsec:community}, respectively.
To this end, we discuss the challenges that hinder industry participation, real-world deployments, and market penetration of research results from a practitioner's point of view.
Afterward, based on our experiences, we report on essential requirements for such interdisciplinary, collaborative research that directly stem from the scientific research community.

\subsection{The Potential of Research Data Management}
\label{subsec:rdm}

First, in Section~\ref{subsubsec:rdmbackground}, we give a background on RDM.
Subsequently, in Section~\ref{subsubsec:rdmmatching}, we detail how to integrate the phases of RDM into our previously proposed process cycle, before discussing the benefits of RDM for (our) cybersecurity research in Section~\ref{subsubsec:rdmpossibilities}.

\subsubsection{A Primer on RDM}
\label{subsubsec:rdmbackground}
The main goal of RDM is to facilitate data handling and sharing within a research project.
Extended to a broader scope, it can also be understood to improve reusability across research projects, between institutes, and over longer time periods.
Research data can include very different kinds of research artifacts, including data in the ``traditional'' sense, \ie databases, data sets, or binary files, possibly comprising (raw) input data, intermediate results, and final results, but also development artifacts, such as code, models, and experiment setups.
Consequently, RDM has to cope with different kinds of information and cover the whole lifespan of data, possibly even beyond project completion.

As a result, RDM is frequently modeled as a lifecycle~\cite{Jones2011Research360,Ball2012Review}.
Accordingly, we illustrate a typical representation of the RDM lifecycle in Figure~\ref{fig:RDMlifecycle}.
In particular, this lifecycle consists of six phases:

\begin{figure}[t]
  \center
  \includegraphics{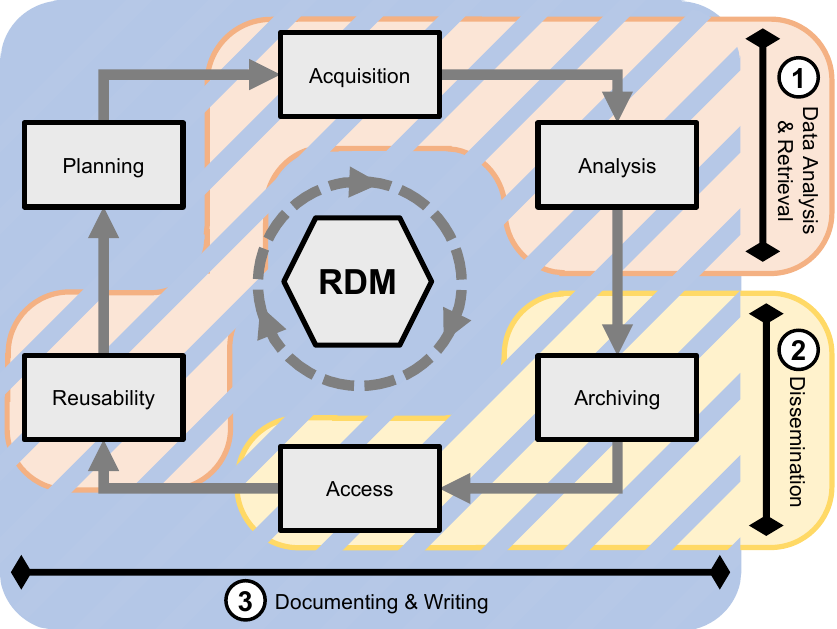}
  \caption{%
    The RDM lifecycle consists of six phases.
    To improve the data handling in our process cycle, we can successfully map the RDM lifecycle phases to it (labeled by \internalstep{1} to \internalstep{3}).
  }
  \label{fig:RDMlifecycle}
\end{figure}

\textbf{Planning.}
The planning step begins at the start of a project or even prior to it.
In this context, research data management plans (RDMPs)~\cite{Hollesetal2018Graduate,Michener2015Ten,Jones2011How} are frequently employed and kept as an evolving research data documentation for each project.
They cover all relevant aspects of data administration, collection, and use within the project.
The respective documentation is updated throughout the complete project and data lifespans.

\textbf{Acquisition.}
Acquiring the relevant data is the first practical step of RDM.
For example, this phase can include the independent generation of data, gathering existing data, or creating metadata.

\textbf{Analysis.}
After obtaining the data, it serves as a source of information throughout the research project.
As such, it can, \eg be used for analysis and interpretation, exchanged with or updated by newer data, or provide the foundation for prototype development.

\textbf{Archiving.}
To subsequently preserve and prove the validity of the research results, the data is archived for long-term re-evaluation and reuse.
As part of this step, to ensure verifiability, data can be migrated to suitable (and reliable) storage mediums and formats.

\textbf{Access.}
An essential goal of RDM is to also grant access to data to other researchers.
Hence, a mechanism to enable data sharing, data distribution, and data publishing must be implemented.

\textbf{Reusability.}
Finally, as the last phase, data should be reusable in other research projects, either by the same organization or other stakeholders.
Thus, this step covers everything related to reusing data as well as reinterpreting data in different contexts.

Next, we map this RDM data lifecycle to our process cycle.

\subsubsection{Introducing RDM to our Process Cycle}
\label{subsubsec:rdmmatching}
Keeping the process cycle (cf.\ Section~\ref{sec:design}) and the RDM lifecycle in mind, we can match the steps from the process cycle to the RDM lifecycle.
As a result, we identify steps where the process cycle can potentially benefit from practices and techniques covered by the RDM lifecycle (Figure~\ref{fig:RDMlifecycle}):

\textbf{\step{1}.} The \hyperref[subsubsec:data]{\emph{Data Analysis \& Retrieval}} step of the process cycle is closely related to the phases \emph{Acquisition}, \emph{Analysis}, and \emph{Reusability} in the RDM lifecycle.
For \emph{Reusability}, we consider the use of older, existing data in a new research project or novel real-world use case.

\textbf{\step{2}.} The \hyperref[subsubsec:dissemination]{\emph{Dissemination}} step of the process cycle addresses the same tasks and issues as the phases \emph{Archiving}, \emph{Access}, and \emph{Reusability} in the RDM lifecycle, \ie it is a fit for RDM practices.

\textbf{\step{3}.} While we consider both reporting and writing (cf.\ \hyperref[subsubsec:writing]{\emph{Reporting \& Writing}}) to be essential tasks in the process cycle, in RDM, the emphasis is rather put on documentation than on reporting.
The task of documentation is present in multiple phases of the RDM lifecycle as all data use is explicitly documented in the beginning (\eg using a RDMP), and continuous planning regarding data and its handling are subsequent fundamental aspects of RDM.

To be precise, the whole RDM lifecycle revolves around organizing the planning and documentation of research data.
The single most important element for the documentation of research data is the RDMP, which aims to break the boundaries of exploring and to ease using other researchers' data without a lot of (repetitive) forth-and-back communication overhead.
The RDM lifecycle, as well as the RDMP, both accompany researchers during the whole project duration.
Hence, RDM is closely related to the \hyperref[subsubsec:writing]{\emph{Reporting \& Writing}} step in our process cycle, which could simultaneously lift the reporting task to a more universally beneficial documenting.

Thus, while RDM is not limited to a single step in our process cycle, we would effectively transition to a \emph{Documenting \& Writing} phase in our cycle to fully address the ideas and needs of RDM.

\subsubsection{Prospects of RDM for (our) Collaborations}
\label{subsubsec:rdmpossibilities}

Considering RDM when working on interdisciplinary research projects at the intersection of cybersecurity and industrial applications promises notable advances in different directions, as we highlight in the following.

\textbf{Impact on Artifacts.}
Enabling collaborations is a central aspect of RDM.
Consequently, RDM constitutes one catalyst for improving the overall state of research by first, giving guidelines on how to organize data and data sharing, and second, by enabling people to reuse each other's results.
The latter further facilitates incremental research, \ie work that directly builds upon or improves the work of others.
In this regard, a noticeable trend is that an increasing number of papers no longer merely presents selected research results, but researchers instead also open-source their research data and/or software artifacts separately.
Likewise, publishers also encourage researchers to include their datasets for the respective work.

Unfortunately, as identified in Section~\ref{subsec:artifacts}, existing incentives for sharing data and prototypes are still often insufficient, and current artifact evaluations have limited expressiveness, thus hindering reusing results and knowledge of others.
However, RDM is also closely linked to the ``culture of researchers'' regarding how research, research continuation, and the sharing of research results are handled.
A growing trend towards the application of RDM could thus change the mindset of researchers and increase the openness for sharing knowledge, data, and code.
Ideally, this change could also impact the publication expectations (cf.\ Section~\ref{subsec:artifacts}) and maybe even encourage the reporting of negative findings (cf.\ Section~\ref{subsubsec:dissemination}).

\textbf{Data Interpretability.}
When working with external data sources, researchers often face the challenge of interpreting data with lacking documentation, missing meta information, or no supporting visualization.
Given that a proper comprehension of the semantic is essential to correctly reuse data, ensuring that everyone interprets the data in the same way is crucial.
Metadata and ontologies, which are essential components of RDM, hereby offer value as they help to bridge wordings, notations, and terminologies between domains.

\textbf{Usage Control.}
Recent advances, such as the International Data Spaces (IDS)~\cite{Baderetal2020The}, also rely on ontologies to improve data reusability.
By offering a secure and reliable data usage control for all stakeholders, the IDS further addresses the issues of privacy, confidentiality, and data sovereignty that still hinder a full implementation of the \emph{Access} and \emph{Archiving} phases of RDM in many organizations.

Especially when considering conservative stakeholders, a standardized (and accepted) approach based on the best practices of RDM and RDMP might help to improve the acceptance of data sharing.
Due to a strict documentation of all steps involved in the data sharing and a better understanding of the implications, this practice might even convince cautious stakeholders to participate.

By design, RDM promises to offer these guarantees:
First, RDM provides researchers with precise guidelines.
Second, a goal of RDM is to propose default tools that ease adhering to these guidelines.
Moreover, RDM also facilitates appropriate and long-lived content and data preservation, an aspect that is beneficial to achieve both (long-term) verifiability and reliability in research environments.

\textbf{Open Issues.}
Unfortunately, RDM is still in its infancy.
In practice, its envisioned tools are neither widely spread, nor yet fully standardized, and often still under development.
To improve this unsatisfactory state, the RDM community is also interested in real-world user stories to adapt their recommended practices accordingly, even though their guidelines capture today's (well-researched) best practices.
Thus, an exchange of experiences is already helpful at this point to evolve RDM on the one hand and our interdisciplinary cybersecurity research methodology on the other hand.

Overall, we conclude that improving our process cycle using practices from RDM to also account for concerns of conservative stakeholders is a viable and sensible aspect. 

\subsection{Driving the Willingness of Real-World Use}
\label{subsec:deployments}

Given that the adoption of developed solutions in industry is an important aspect of rating the success of collaborations for practical use, we now discuss this aspect from a practitioner's point of view.

\textbf{Implementing Research Results.}
Research projects like our cluster of excellence offer the unique opportunity to transfer successful results from academia to industry by enabling long-term sharing of experimental data and executable models.
However, the data and models are often highly sensitive, especially when they are developed together with practitioners from the industry.
Therefore, all solutions must consider cybersecurity and privacy needs at an early stage of the project to develop viable approaches for real-world deployment.
The recent privacy-by-design paradigm~\cite{Cavoukian2009Privacy,Gursesetal2011Engineering} promises to particularly address this aspect.
Thus, conforming solutions could enable data sharing and knowledge transfers within the industry (and possibly even with academia).
The efforts of working on these solutions must be synchronized with the strategy of ``balancing openness and knowledge transfer'', which is relevant in science communication, technology transfer, research data management, and further publication mediums.
In particular, the grade of openness is directly driven by the community, \ie the specific type of product that is produced and the company's position in a supply chain (\eg OEM or supplier)~\cite{Schuhetal2017Product}.

\textbf{Diverse Industrial Needs.}
For the needs of the producing industry, we have to distinguish the different requirements and challenges of each domain.
Independent of any approach, all producers strive to find an optimum within the polylemma of production, \ie scale or scope, value-oriented or planning-oriented.
Relying on data sharing and knowledge transfers inheres the potential of solving this polylemma~\cite{Brecheretal2017Integrative}.
The value and criticality of manufacturing data for a manufacturer of blisks (blade integrated disc), which produces critical components of an aero-engine part is different from that of cheap screws for furniture.
Although the abstraction and modeling process described in Figure~\ref{fig:processcycle} might be similar, a strict research focus and appropriate time and resource planning are necessary for applied sciences.
Consequently, this aspect mainly influences the cycle's bootstrapping phase (cf.\ Section~\ref{subsubsec:bootstrapping}).

Next, we exemplarily discuss an aero-engine manufacturer.

\textbf{Driving the Willingness of Big-Tech Industries.}
Within Big Tech, the aircraft industry is one of the most considerable domains:
The structural parts of an aircraft are specified by form tolerances of only a few micrometers.
During production, relevant context information (\eg form tolerances) is extracted and connected to reduced-order models.
IT-driven advances, such as CPS and the IIoT, are expected to refine related production processes by combining various data sources (even across domains), \ie knowledge is exchanged globally to improve the decision-making in the industry.

For example, a real-time machining simulation to support the production, which is derived from models of multiple stakeholders, might be more accurate in comparison to the status quo.
Likewise, a parallel material removal simulation allows for predicting current process forces and resulting deflections.
Enhancing the accuracy of these predictions using external data might increase the producer's productivity as the reliability and correctness of decisions to, \eg rework or scrap the part are improved.
Utilizing all this information from all data sources might allow for changes across machines, production technologies, and even supply chains.

Already today, companies would like to predict the part quality during milling, on the one hand, to ensure the quality and to collect the data with sufficient boundary conditions.
On the other hand, process data is mostly considered confidential (\eg the form tolerances or the machining parameters), and thus, information is only shared if required by law (\eg to aviation authorities).
An example that strives to overcome these isolated data silos is the Data4Safety initiative~\cite{European-Union-Aviation-Safety-Agency2017Data4Safety:}, which collects information on boundary conditions.

As shown with these examples, Big Tech pursues an optimum between data privacy and openness w.r.t.\ their productivity.
This focus already has an impact on legislation and deployed approaches (\eg~\cite{Mandollaetal2019Building}).
Hence, research that targets novel use cases and applications, such as the ones enabled by cybersecurity, also has to primarily consider this optimum to achieve real-world deployments.

\textbf{Raising Awareness.}
This philosophy must be considered at all times by stakeholders from industry, academia, and government.
Joint projects form the contextual framework for (successful) long-term collaborations, which eventually also serve as incubators for additional research proposals.
To ensure their success, data privacy and security should be approached and eventually integrated in a similar way as approaches for business models in the context of digitization in the past~\cite{Burmeisteretal2016Business}, \ie an incremental transformation will be successful when gradually implementing novel (and secure) approaches that stem from academic research in productive use.

From a practitioner's point of view, increasing the willingness of real-world use and raising the awareness of new approaches can only succeed with joint efforts that gradually push the novelties of interdisciplinary cybersecurity research into these (traditionally) conservative domains.
For an opposing view, we next discuss the constraints imposed by academia for this challenging endeavor.

\subsection{Addressing the Needs of Academic Research}
\label{subsec:community}

In addition to the expectations formulated by industry, the scientific community also has a significant impact on the interdisciplinary work that we focus our collaborative research on.
In the following, we discuss the resulting consequences and open issues.

\textbf{Suitable Target Community.}
A crucial aspect of the publishing process is to select a fitting research community (cf.\ Section~\ref{subsubsec:writing}).
Due to our focus on cybersecurity research for industrial applications, our work does not match the traditional expectations of fundamental work in computer science.
Thus, we explicitly have to select venues that also accept work on practical applications.

Irrespective of the small number of calls for papers that focus on this novel research area, we noticed that interdisciplinary work can negatively influence the chances of publication.
On the one hand, program committee members and (external) reviewers might not be able to correctly understand the real-world impact given their lack of required domain knowledge, an issue that we raised several times when introducing our process cycle (cf.\ Section~\ref{sec:design}).
This situation is expected as computer scientists usually do not have a degree in, for example, mechanical engineering.
On the other hand, reviewers still require a sound story to accept such practical work for publication.
Unfortunately, in most cases, the presented work should be tied to one or multiple use cases that are already deployed in the real world.
In our view, this demand corresponds to a classic ``chicken or the egg'' situation as unpublished work cannot receive a large market penetration (cf.\ Section~\ref{subsec:deployments}).
Hence, it hinders the research progress and challenges the work's practical acceptance.

\textbf{Evaluation Concerns.}
A key criterion of every paper is the soundness and relevance of the evaluation.
As such, we highly recommend the use of real-world use case data (cf.\ Section~\ref{subsubsec:evaluation}).
Unfortunately, sufficient data might not be available or might not be shared (\eg due to confidentiality).
In these situations, using artificial use case data is an alternative.
However, paired with the practical focus of our considered interdisciplinary work, the acceptance of this data is low.
Here, a concern is that the validity of the used data cannot be attested, which impacts the presented results.

When initially submitting our work on practical data interoperability (later published at \ref{paper:btw}), we aimed for a call for papers that explicitly accepted practical work.
Our paper was rejected from the in-use track as our work was not yet (widely) deployed.
Certainly, most of our research is on a conceptual level that focuses on proof of concepts and prototypes rather than demonstrators.
However, this situation highlights that such interdisciplinary research lies somewhere in between fundamental and practical research, effectively within a gap that only attracts interest from a few venues.

\textbf{Focus of Dissemination.}
As discussed in Section~\ref{subsubsec:dissemination}, the expectations between industry and academia often differ concerning the readiness level of the developed approach.
From a scientific point of view, further work to build upon prototypes only offers little added contribution.
Thus, the impact of such efforts on researchers is severely limited, even when considering the interdisciplinary scope of said work.
Efforts to ``finish'' prototypes are not even remotely linked to the usually preferred fundamental research.

From this dilemma, we concluded that both researchers and practitioners should (individually) address their specific needs:
By focusing on the contributions that are relevant to the respective domain, the chances to publish the developed approach improve.
On the one hand, researchers should try to highlight the universality of their work to underline the importance and relevance for the cybersecurity research community.
On the other hand, practitioners can subsequently report on the impact a deployment of a novel interdisciplinary solution had on their use case, \ie focus on domain specifics rather than the details of the underlying security and privacy concept.
Thus, the involved parties can simultaneously improve the awareness for this novel research area in both domains.

At this point, we advise to caution as repeated publications of results might contradict \new{the widely-established} standards of good scientific conduct~\cite{Deutsche-Forschungsgemeinschaft2019Guidelines}.

\section{Conclusion}
\label{sec:conclusion}

Based on our experience during several collaborative research projects, aided by our cluster of excellence Internet of Production, we derived a methodology for conducting practical research on realizing novel forms of industrial applications and collaboration through recent advances in cybersecurity and privacy research.
This abstract process cycle is meant to support researchers in this emerging research area, which is significantly evolving due to recent trends in digitization, such as CPS and the IoT.
Illustrated with our research experiences, we are able to provide a realistic and plausible overview of common challenges and pitfalls.

Overall, we ease the challenges of and reservations against interdisciplinary collaborations and hope to contribute to bootstrapping additional research in this area---an area where a lot of untapped potential remains.
Despite these advances, we still identify that some friction at the intersection between academia and industry remains, which should be addressed in the future by gradually converging the respective mindsets.
We believe that the practical implementation of research data management into our process cycle can help to overcome some concerns that are connected to confidentiality and privacy, a major obstacle in conducting real-world applicable evaluations and achieving artifacts reusability (across stakeholders).

We look forward to all upcoming research in this \new{developing} area and expect first deployed solutions in productive use and development \new{with significant real-world impact} soon.


\section*{Acknowledgment}

We would like to thank all of our incredibly supporting collaborators, the reviewers, and David Balenson for their fruitful comments and invaluable support.
Funded by the Deutsche Forschungsgemeinschaft (DFG, German Research Foundation) under Germany's Excellence Strategy -- EXC-2023 Internet of Production -- 390621612.



%

\bibliographystyle{IEEEtranS}
\bibliography{paper}

\begin{thebibliography}{100}
\providecommand{\url}[1]{#1}
\csname url@samestyle\endcsname
\providecommand{\newblock}{\relax}
\providecommand{\bibinfo}[2]{#2}
\providecommand{\BIBentrySTDinterwordspacing}{\spaceskip=0pt\relax}
\providecommand{\BIBentryALTinterwordstretchfactor}{4}
\providecommand{\BIBentryALTinterwordspacing}{\spaceskip=\fontdimen2\font plus
\BIBentryALTinterwordstretchfactor\fontdimen3\font minus
  \fontdimen4\font\relax}
\providecommand{\BIBforeignlanguage}[2]{{%
\expandafter\ifx\csname l@#1\endcsname\relax
\typeout{** WARNING: IEEEtranS.bst: No hyphenation pattern has been}%
\typeout{** loaded for the language `#1'. Using the pattern for}%
\typeout{** the default language instead.}%
\else
\language=\csname l@#1\endcsname
\fi
#2}}
\providecommand{\BIBdecl}{\relax}
\BIBdecl

\bibitem{ACM2020Artifact}
{ACM}, ``{Artifact Review and Badging - Current},''
  \url{https://www.acm.org/publications/policies/artifact-review-and-badging-current},
  2020.

\bibitem{Anietal2021Design}
U.~P.~D. Ani, J.~M. Watson, B.~Green, B.~Craggs, and J.~R.~C. Nurse, ``{Design
  Considerations for Building Credible Security Testbeds: Perspectives from
  Industrial Control System Use Cases},'' \emph{Journal of Cyber Security
  Technology}, vol.~5, no.~2, 2021.

\bibitem{Arpetal2022Dos}
D.~Arp, E.~Quiring, F.~Pendlebury, A.~Warnecke, F.~Pierazzi, C.~Wressnegger,
  L.~Cavallaro, and K.~Rieck, ``{Dos and Don'ts of Machine Learning in Computer
  Security},'' in \emph{Proceedings of the 31st USENIX Security Symposium (SEC
  '22)}.\hskip 1em plus 0.5em minus 0.4em\relax USENIX Association, 2022.

\bibitem{Ayetal2019Kernel}
M.~Ay, D.~Stenger, M.~Schwenzer, D.~Abel, and T.~Bergs, ``{Kernel Selection for
  Support Vector Machines for System Identification of a CNC Machining
  Center},'' \emph{IFAC-PapersOnLine}, vol.~52, no.~29, pp. 192--198, 2019.

\bibitem{Baderetal2021Blockchain-Based}
L.~Bader, J.~Pennekamp, R.~Matzutt, D.~Hedderich, M.~Kowalski, V.~L{\"u}cken,
  and K.~Wehrle, ``{Blockchain-Based Privacy Preservation for Supply Chains
  Supporting Lightweight Multi-Hop Information Accountability},''
  \emph{Information Processing {\&} Management}, vol.~58, no.~3, 2021.

\bibitem{Baderetal2020The}
S.~Bader, J.~Pullmann, C.~Mader, S.~Tramp \emph{et~al.}, ``{The International
  Data Spaces Information Model - An Ontology for Sovereign Exchange of Digital
  Content},'' in \emph{Proceedings of the 19th International Semantic Web
  Conference on The Semantic Web (ISWC '20)}, vol. 12507.\hskip 1em plus 0.5em
  minus 0.4em\relax Springer, 2020, pp. 176--192.

\bibitem{Balensonetal2015Cybersecurity}
D.~Balenson, L.~Tinnel, and T.~Benzel, ``{Cybersecurity Experimentation of the
  Future (CEF): Catalyzing a New Generation of Experimental Cybersecurity
  Research},'' SRI International, Tech. Rep., 2015.

\bibitem{Ball2012Review}
A.~Ball, ``{Review of Data Management Lifecycle Models},'' University of Bath,
  Tech. Rep., 2012.

\bibitem{Ballotetal2014The}
E.~Ballot, B.~Montreuil, and R.~Meller, \emph{{The Physical Internet}}.\hskip
  1em plus 0.5em minus 0.4em\relax La Documentation Fran{\c c}aise, 2014.

\bibitem{Beckeretal2021A}
F.~Becker, P.~Bibow, M.~Dalibor, A.~Gannouni \emph{et~al.}, ``{A Conceptual
  Model for Digital Shadows in Industry and its Application},'' in
  \emph{Proceedings of the 40th International Conference on Conceptual Modeling
  (ER'21)}, vol. 13011.\hskip 1em plus 0.5em minus 0.4em\relax Springer, 2021,
  pp. 271--281.

\bibitem{Beheryetal2020Action}
M.~Behery, M.~Tschesche, F.~Rudolph, G.~Hirt, and G.~Lakemeyer, ``{Action
  Discretization for Robot Arm Teleoperation in Open-Die Forging},'' in
  \emph{Proceedings of the 2020 IEEE International Conference on Systems, Man,
  and Cybernetics (SMC '20)}.\hskip 1em plus 0.5em minus 0.4em\relax IEEE,
  2020, pp. 2100--2105.

\bibitem{Ben-Salemetal2011On}
M.~Ben~Salem and S.~Stolfo, ``{On the Design and Execution of Cyber-Security
  User Studies: Methodology, Challenges, and Lessons Learned},'' in
  \emph{Proceedings of the 4th Workshop on Cyber Security Experimentation and
  Test (CSET '11)}.\hskip 1em plus 0.5em minus 0.4em\relax USENIX Association,
  2011.

\bibitem{Bergeretal2018SIGPLAN}
\BIBentryALTinterwordspacing
E.~D. Berger, S.~M. Blackburn, M.~Hauswirth, and M.~Hicks, ``{SIGPLAN Empirical
  Evaluation Checklist},'' ACM SIGPLAN, Tech. Rep., 2018. [Online]. Available:
  \url{https://www.sigplan.org/Resources/EmpiricalEvaluation/}
\BIBentrySTDinterwordspacing

\bibitem{Bibowetal2020Model-Driven}
P.~Bibow, M.~Dalibor, C.~Hopmann, B.~Mainz, B.~Rumpe, D.~Schmalzing,
  M.~Schmitz, and A.~Wortmann, ``{Model-Driven Development of a Digital Twin
  for Injection Molding},'' in \emph{Proceedings of the 32nd International
  Conference on Advanced Information Systems Engineering (CAiSE '20)}, vol.
  12127.\hskip 1em plus 0.5em minus 0.4em\relax Springer, 2020, pp. 85--100.

\bibitem{Bodenbenneretal2021FAIR}
M.~Bodenbenner, B.~Montavon, and R.~H. Schmitt, ``{FAIR sensor services -
  Towards sustainable sensor data management},'' \emph{Measurement: Sensors},
  vol.~18, 2021.

\bibitem{Bodenbenneretal2020Domain-Specific}
M.~Bodenbenner, M.~P. Sanders, B.~Montavon, and R.~H. Schmitt,
  ``{Domain-Specific Language for Sensors in the Internet of Production},'' in
  \emph{Proceedings of the 10th Congress of the German Academic Association for
  Production Technology (WGP '20)}, vol.~20.\hskip 1em plus 0.5em minus
  0.4em\relax Springer, 2020, pp. 448--456.

\bibitem{Braunetal2020An}
S.~Braun, I.~Koren, M.~Van~Dyck, and M.~Jarke, ``{An Agricultural Data Platform
  iStar Model},'' in \emph{Proceedings of the 13th International iStar Workshop
  (iStar '20)}, vol. 2641.\hskip 1em plus 0.5em minus 0.4em\relax CEUR Workshop
  Proceedings, 2020, pp. 19--24.

\bibitem{Brauneretal2022A}
P.~Brauner, M.~Dalibor, M.~Jarke, I.~Kunze \emph{et~al.}, ``{A Computer Science
  Perspective on Digital Transformation in Production},'' \emph{ACM
  Transactions on Internet of Things}, 2022.

\bibitem{Brecheretal2021Gaining}
C.~Brecher, M.~Buchsbaum, A.~M{\"u}ller, K.~Schilling, M.~Obdenbusch,
  S.~Staudacher, and M.~C. Albasatineh, ``{Gaining IIoT insights by leveraging
  ontology-based modelling of raw data and Digital Shadows},'' in
  \emph{Proceedings of the 2021 4th IEEE International Conference on Industrial
  Cyber-Physical Systems (ICPS '21)}.\hskip 1em plus 0.5em minus 0.4em\relax
  IEEE, 2021, pp. 231--236.

\bibitem{Brecheretal2017Integrative}
C.~Brecher, D.~{\"O}zdemir, and A.~R. Weber, \emph{{Integrative Production
  Technology---Theory and Applications}}.\hskip 1em plus 0.5em minus
  0.4em\relax Springer, 2017, pp. 1--17.

\bibitem{Brillowskietal2021Know-How}
F.~Brillowski, H.~Dammers, H.~Koch, K.~M{\"u}ller, L.~Reinsch, and C.~Greb,
  ``{Know-How Transfer and Production Support Systems to Cultivate the Internet
  of Production Within the Textile Industry},'' in \emph{Proceedings of the 4th
  International Conference on Intelligent Human Systems Integration (IHSI
  '21)}, vol. 1322.\hskip 1em plus 0.5em minus 0.4em\relax Springer, 2021, pp.
  309--315.

\bibitem{Brockhoffetal2021Process}
T.~Brockhoff, M.~Heithoff, I.~Koren, J.~Michael \emph{et~al.}, ``{Process
  Prediction with Digital Twins},'' in \emph{Companion Proceedings of the
  ACM/IEEE 24th International Conference on Model Driven Engineering Languages
  and Systems (MODELS-C '21)}.\hskip 1em plus 0.5em minus 0.4em\relax IEEE,
  2021.

\bibitem{Buckhorstetal2021Holarchy}
A.~F. Buckhorst, B.~Montavon, D.~Wolfschl{\"a}ger, M.~Buchsbaum \emph{et~al.},
  ``{Holarchy for Line-less Mobile Assembly Systems Operation in the Context of
  the Internet of Production},'' \emph{Procedia CIRP}, vol.~99, pp. 448--453,
  2021, proceedings of the 14th CIRP Conference on Intelligent Computation in
  Manufacturing Engineering (ICME '20).

\bibitem{Burmeisteretal2016Business}
C.~Burmeister, D.~L{\"u}ttgens, and F.~T. Piller, ``{Business Model Innovation
  for Industrie 4.0: Why the ``Industrial Internet'' Mandates a New Perspective
  on Innovation},'' \emph{Die Unternehmung}, vol.~70, no.~2, pp. 124--152,
  2016.

\bibitem{Buschmannetal2021Data-driven}
D.~Buschmann, C.~Enslin, H.~Elser, D.~L{\"u}tticke, and R.~H. Schmitt,
  ``{Data-driven decision support for process quality improvements},''
  \emph{Procedia CIRP}, vol.~99, pp. 313--318, 2021.

\bibitem{Carrolletal2012Realizing}
T.~E. Carroll, D.~Manz, T.~Edgar, and F.~L. Greitzer, ``{Realizing scientific
  methods for cyber security},'' in \emph{Proceedings of the 2012 Workshop on
  Learning from Authoritative Security Experiment Results (LASER '12)}.\hskip
  1em plus 0.5em minus 0.4em\relax ACM, 2012, pp. 19--24.

\bibitem{Cavoukian2009Privacy}
A.~Cavoukian, ``{Privacy by Design},'' Information and Privacy Commissioner of
  Ontario, Tech. Rep., 2009.

\bibitem{Collbergetal2015A}
C.~Collberg and T.~Proebsting, ``{A Catalog of Research Artifacts for Computer
  Science},'' \url{http://www.findresearch.org}, 2015.

\bibitem{COMSYS2020Privacy-Preserving}
{COMSYS}, ``{Privacy-Preserving Production Process Parameter Exchange},''
  \url{https://github.com/COMSYS/parameter-exchange}, 2020.

\bibitem{Contietal2021A}
M.~Conti, D.~Donadel, and F.~Turrin, ``{A Survey on Industrial Control System
  Testbeds and Datasets for Security Research},'' arXiv:2102.05631, 2021.

\bibitem{Crameretal2021Towards}
S.~Cramer, M.~Hoffmann, P.~Schlegel, M.~Kemmerling, and R.~H. Schmitt,
  ``{Towards a flexible process-independent meta-model for production data},''
  \emph{Procedia CIRP}, vol.~99, pp. 586--591, 2021.

\bibitem{Crusselletal2019Lessons}
J.~Crussell, T.~M. Kroeger, D.~Kavaler, A.~Brown, and C.~Phillips, ``{Lessons
  Learned from 10k Experiments to Compare Virtual and Physical Testbeds},'' in
  \emph{Proceedings of the 12th USENIX Workshop on Cyber Security
  Experimentation and Test (CSET '19)}.\hskip 1em plus 0.5em minus 0.4em\relax
  USENIX Association, 2019.

\bibitem{Dahlmannsetal2019Privacy-Preserving}
M.~Dahlmanns, C.~Dax, R.~Matzutt, J.~Pennekamp, J.~Hiller, and K.~Wehrle,
  ``{Privacy-Preserving Remote Knowledge System},'' in \emph{Proceedings of the
  2019 IEEE 27th International Conference on Network Protocols (ICNP
  '19)}.\hskip 1em plus 0.5em minus 0.4em\relax IEEE, 2019.

\bibitem{Dahlmannsetal2020Easing}
M.~Dahlmanns, J.~Lohm{\"o}ller, I.~B. Fink, J.~Pennekamp, K.~Wehrle, and
  M.~Henze, ``{Easing the Conscience with OPC UA: An Internet-Wide Study on
  Insecure Deployments},'' in \emph{Proceedings of the ACM Internet Measurement
  Conference (IMC '20)}.\hskip 1em plus 0.5em minus 0.4em\relax ACM, 2020, pp.
  101--110.

\bibitem{Dahlmannsetal2022Missed}
M.~Dahlmanns, J.~Lohm{\"o}ller, J.~Pennekamp, J.~Bodenhausen, K.~Wehrle, and
  M.~Henze, ``{Missed Opportunities: Measuring the Untapped TLS Support in the
  Industrial Internet of Things},'' in \emph{Proceedings of the 17th ACM ASIA
  Conference on Computer and Communications Security (ASIACCS '22)}.\hskip 1em
  plus 0.5em minus 0.4em\relax ACM, 2022.

\bibitem{Dahlmannsetal2021Transparent}
M.~Dahlmanns, J.~Pennekamp, I.~B. Fink, B.~Schoolmann, K.~Wehrle, and M.~Henze,
  ``{Transparent End-to-End Security for Publish/Subscribe Communication in
  Cyber-Physical Systems},'' in \emph{Proceedings of the 1st ACM Workshop on
  Secure and Trustworthy Cyber-Physical Systems (SaT-CPS '21)}.\hskip 1em plus
  0.5em minus 0.4em\relax ACM, 2021, pp. 78--87.

\bibitem{Deutsche-Forschungsgemeinschaft2019Guidelines}
{Deutsche Forschungsgemeinschaft}, ``{Guidelines for Safeguarding Good Research
  Practice},'' German Research Foundation, Code of Conduct, 2019.

\bibitem{Dumitrasetal2011Experimental}
T.~Dumitras and I.~Neamtiu, ``{Experimental Challenges in Cyber Security: A
  Story of Provenance and Lineage for Malware},'' in \emph{Proceedings of the
  4th Workshop on Cyber Security Experimentation and Test (CSET '11)}.\hskip
  1em plus 0.5em minus 0.4em\relax USENIX Association, 2011.

\bibitem{Dykstraetal2019Lessons}
J.~Dykstra, M.~Fante, P.~Donahue, D.~Varva, L.~Wilk, and A.~Johnson, ``{Lessons
  from Using the I-Corps Methodology to Understand Cyber Threat Intelligence
  Sharing},'' in \emph{Proceedings of the 12th USENIX Workshop on Cyber
  Security Experimentation and Test (CSET '19)}.\hskip 1em plus 0.5em minus
  0.4em\relax USENIX Association, 2019.

\bibitem{Dykstraetal2018Cyber}
J.~Dykstra and C.~L. Paul, ``{Cyber Operations Stress Survey (COSS): Studying
  fatigue, frustration, and cognitive workload in cybersecurity operations},''
  in \emph{Proceedings of the 11th USENIX Workshop on Cyber Security
  Experimentation and Test (CSET '18)}.\hskip 1em plus 0.5em minus 0.4em\relax
  USENIX Association, 2018.

\bibitem{Elseretal2021Potentials}
H.~Elser, P.~Jongebloed, D.~Buschmann, M.~Ellerich, and R.~H. Schmitt,
  ``{Potentials of Bluetooth Low Energy Beacons for order tracing in single and
  small batch production},'' \emph{Procedia CIRP}, vol.~97, pp. 202--210, 2021,
  proceedings of the 8th CIRP Conference of Assembly Technology and Systems
  (CATS '21).

\bibitem{European-Union-Aviation-Safety-Agency2017Data4Safety:}
{European Union Aviation Safety Agency}, ``{Data4Safety: A partnership for a
  data driven aviation safety analysis in Europe},''
  \url{https://www.easa.europa.eu/newsroom-and-events/news/data4safety-partnership-data-driven-aviation-safety-analysis-europe},
  2017 (accessed February 25, 2020).

\bibitem{Evaluate-Collaboratory2010Experimental}
{Evaluate Collaboratory}, ``{Experimental Evaluation of Software and Systems in
  Computer Science},'' \url{http://evaluate.inf.usi.ch/}, 2010.

\bibitem{Garneauetal2016Results}
C.~J. Garneau, R.~F. Erbacher, R.~E. Etoty, and S.~E. Hutchinson, ``{Results
  and Lessons Learned from a User Study of Display Effectiveness with
  Experienced Cyber Security Network Analysts},'' in \emph{Proceedings of the
  2016 Workshop on Learning from Authoritative Security Experiment Results
  (LASER '16)}.\hskip 1em plus 0.5em minus 0.4em\relax USENIX Association,
  2016, pp. 33--42.

\bibitem{Glebkeetal2019A}
R.~Glebke, M.~Henze, K.~Wehrle, P.~Niemietz, D.~Trauth, P.~Mattfeld, and
  T.~Bergs, ``{A Case for Integrated Data Processing in Large-Scale
  Cyber-Physical Systems},'' in \emph{Proceedings of the 52nd Hawaii
  International Conference on System Sciences (HICSS '19)}.\hskip 1em plus
  0.5em minus 0.4em\relax AIS, 2019, pp. 7252--7261.

\bibitem{Glebkeetal2019Towards}
R.~Glebke, J.~Krude, I.~Kunze, J.~R{\"u}th, F.~Senger, and K.~Wehrle,
  ``{Towards Executing Computer Vision Functionality on Programmable Network
  Devices},'' in \emph{Proceedings of the 1st ACM CoNEXT Workshop on Emerging
  in-Network Computing Paradigms (ENCP '19)}.\hskip 1em plus 0.5em minus
  0.4em\relax ACM, 2019, pp. 15--20.

\bibitem{Gleimetal2020Timestamped}
L.~Gleim and S.~Decker, ``{Timestamped URLs as Persistent Identifiers},'' in
  \emph{Proceedings of the 6th Workshop on Managing the Evolution and
  Preservation of the Data Web (MEPDaW '20)}, vol. 2821.\hskip 1em plus 0.5em
  minus 0.4em\relax CEUR Workshop Proceedings, 2020, pp. 11--16.

\bibitem{Gleimetal2020FactDAG:}
L.~Gleim, J.~Pennekamp, M.~Liebenberg, M.~Buchsbaum \emph{et~al.}, ``{FactDAG:
  Formalizing Data Interoperability in an Internet of Production},'' \emph{IEEE
  Internet of Things Journal}, vol.~7, no.~4, pp. 3243--3253, 2020.

\bibitem{Gleimetal2021FactStack:}
L.~Gleim, J.~Pennekamp, L.~Tirpitz, S.~Welten, F.~Brillowski, and S.~Decker,
  ``{FactStack: Interoperable Data Management and Preservation for the Web and
  Industry 4.0},'' in \emph{Proceedings of the 19th Symposium for Database
  Systems for Business, Technology and Web (BTW '21)}, vol. P-312.\hskip 1em
  plus 0.5em minus 0.4em\relax Gesellschaft f{\"u}r Informatik, 2021, pp.
  371--395.

\bibitem{Gleimetal2020Expressing}
L.~Gleim, L.~Tirpitz, J.~Pennekamp, and S.~Decker, ``{Expressing FactDAG
  Provenance with PROV-O},'' in \emph{Proceedings of the 6th Workshop on
  Managing the Evolution and Preservation of the Data Web (MEPDaW '20)}, vol.
  2821.\hskip 1em plus 0.5em minus 0.4em\relax CEUR Workshop Proceedings, 2020,
  pp. 53--58.

\bibitem{Greenetal2017Pains}
B.~Green, A.~Lee, R.~Antrobus, U.~Roedig, D.~Hutchison, and A.~Rashid,
  ``{Pains, Gains and PLCs: Ten Lessons from Building an Industrial Control
  Systems Testbed for Security Research},'' in \emph{Proceedings of the 10th
  USENIX Workshop on Cyber Security Experimentation and Test (CSET '17)}.\hskip
  1em plus 0.5em minus 0.4em\relax USENIX Association, 2017.

\bibitem{Grochowskietal2019Applying}
M.~Grochowski, S.~Kowalewski, M.~Buchsbaum, and C.~Brecher, ``{Applying Runtime
  Monitoring to the Industrial Internet of Things},'' in \emph{Proceedings of
  the 2019 24th IEEE International Conference on Emerging Technologies and
  Factory Automation (ETFA '19)}.\hskip 1em plus 0.5em minus 0.4em\relax IEEE,
  2019, pp. 348--355.

\bibitem{Gursesetal2011Engineering}
S.~G{\"u}rses, C.~Troncoso, and C.~Diaz, ``{Engineering Privacy by Design},''
  in \emph{Proceedings of the 4th International Conference on Computers,
  Privacy {\&} Data Protection (CPDP '11)}.\hskip 1em plus 0.5em minus
  0.4em\relax CPDP Conferences, 2011.

\bibitem{Henze2020The}
M.~Henze, ``{The Quest for Secure and Privacy-preserving Cloud-based Industrial
  Cooperation},'' in \emph{Proceedings of the 2020 IEEE Conference on
  Communications and Network Security (CNS '20)}.\hskip 1em plus 0.5em minus
  0.4em\relax IEEE, 2020.

\bibitem{Henzeetal2017Network}
M.~Henze, J.~Hiller, R.~Hummen, R.~Matzutt, K.~Wehrle, and J.~H. Ziegeldorf,
  \emph{{Network Security and Privacy for Cyber-Physical Systems}}.\hskip 1em
  plus 0.5em minus 0.4em\relax Wiley, 2017, pp. 25--56.

\bibitem{Hilleretal2018Secure}
J.~Hiller, M.~Henze, M.~Serror, E.~Wagner, J.~N. Richter, and K.~Wehrle,
  ``{Secure Low Latency Communication for Constrained Industrial IoT
  Scenarios},'' in \emph{Proceedings of the 2018 IEEE 43rd Conference on Local
  Computer Networks (LCN '18)}.\hskip 1em plus 0.5em minus 0.4em\relax IEEE,
  2018, pp. 614--622.

\bibitem{Hollesetal2018Graduate}
J.~H. Holles and L.~Schmidt, ``{Graduate Research Data Management Course
  Content: Teaching the Data Management Plan (DMP)},'' in \emph{Proccedings of
  the 126th 2018 ASEE Annual Conference {\&} Exposition}.\hskip 1em plus 0.5em
  minus 0.4em\relax American Society for Engineering Education, 2018.

\bibitem{Intel2020IntelR}
{Intel}, ``{Intel{\textregistered} Tofino{\texttrademark}},''
  \url{https://www.intel.com/content/www/us/en/products/network-io/programmable-ethernet-switch/tofino-series.html},
  2020.

\bibitem{Jarke2020Data}
M.~Jarke, ``{Data Sovereignty and the Internet of Production},'' in
  \emph{Proceedings of the 32nd International Conference on Advanced
  Information Systems Engineering (CAiSE '20)}, vol. 12127.\hskip 1em plus
  0.5em minus 0.4em\relax Springer, 2020, pp. 549--558.

\bibitem{Jones2011Research360}
K.~Jones, ``{Research360 @ Bath - Managing data across the institutional
  research lifecycle},'' in \emph{Proceedings of the 7th International Digital
  Curation Conference (IDCC '11)}.\hskip 1em plus 0.5em minus 0.4em\relax
  Digital Curation Centre, 2011.

\bibitem{Jones2011How}
S.~Jones, ``{How to Develop a Data Management and Sharing Plan},'' Edinburgh:
  Digital Curation Centre, DCC How-to Guides, 2011.

\bibitem{Kirchhofetal2022MontiThings:}
J.~C. Kirchhof, B.~Rumpe, D.~Schmalzing, and A.~Wortmann, ``{MontiThings:
  Model-driven Development and Deployment of Reliable IoT Applications},''
  \emph{Journal of Systems and Software}, vol. 183, 2022.

\bibitem{Kroletal2016Towards}
K.~Krol, J.~M. Spring, S.~Parkin, and M.~A. Sasse, ``{Towards robust
  experimental design for user studies in security and privacy},'' in
  \emph{Proceedings of the 2016 Workshop on Learning from Authoritative
  Security Experiment Results (LASER '16)}.\hskip 1em plus 0.5em minus
  0.4em\relax USENIX Association, 2016, pp. 21--31.

\bibitem{Kunzeetal2021Investigating}
I.~Kunze, R.~Glebke, J.~Scheiper, M.~Bodenbenner, R.~H. Schmitt, and K.~Wehrle,
  ``{Investigating the Applicability of In-Network Computing to Industrial
  Scenarios},'' in \emph{Proceedings of the 2021 4th IEEE International
  Conference on Industrial Cyber-Physical Systems (ICPS '21)}.\hskip 1em plus
  0.5em minus 0.4em\relax IEEE, 2021, pp. 334--340.

\bibitem{Kunzeetal2021Detecting}
I.~Kunze, P.~Niemietz, L.~Tirpitz, R.~Glebke, D.~Trauth, T.~Bergs, and
  K.~Wehrle, ``{Detecting Out-Of-Control Sensor Signals in Sheet Metal Forming
  using In-Network Computing},'' in \emph{Proceedings of the 2021 IEEE 30th
  International Symposium on Industrial Electronics (ISIE '21)}.\hskip 1em plus
  0.5em minus 0.4em\relax IEEE, 2021.

\bibitem{LASER2011The}
{LASER}, ``{The LASER Workshop},'' \url{https://laser-workshop.org/}, 2013.

\bibitem{Levesqueetal2014Computer}
F.~L. L{\'e}vesque and J.~M. Fernandez, ``{Computer Security Clinical Trials:
  Lessons Learned from a 4-month Pilot Study},'' in \emph{Proceedings of the
  7th Workshop on Cyber Security Experimentation and Test (CSET '14)}.\hskip
  1em plus 0.5em minus 0.4em\relax USENIX Association, 2014.

\bibitem{Liebenbergetal2020Information}
M.~Liebenberg and M.~Jarke, ``{Information Systems Engineering with
  DigitalShadows: Concept and Case Studies},'' in \emph{Proceedings of the 32nd
  International Conference on Advanced Information Systems Engineering (CAiSE
  '20)}, vol. 12127.\hskip 1em plus 0.5em minus 0.4em\relax Springer, 2020, pp.
  70--84.

\bibitem{Linetal2020Cyber-Physical}
H.~Lin, B.~Shrestha, and Y.-C. Hu, ``{Cyber-Physical Testbed: Case Study to
  Evaluate Anti-Reconnaissance Approaches on Power Grids' Cyber-Physical
  Infrastructures},'' in \emph{Proceedings of the 2020 Workshop on Learning
  from Authoritative Security Experiment Results (LASER '20)}.\hskip 1em plus
  0.5em minus 0.4em\relax Internet Society, 2020.

\bibitem{Lippetal2020Flexible}
J.~Lipp, M.~Rath, M.~Rudack, U.~Vroomen, and A.~B{\"u}hrig-Polaczek,
  \emph{{Flexible OPC UA Data Load Optimizations on the Edge of
  Production}}.\hskip 1em plus 0.5em minus 0.4em\relax Springer, 2020, pp.
  43--61.

\bibitem{Lippetal2020When}
J.~Lipp, M.~Rudack, U.~Vroomen, and A.~B{\"u}hrig-Polaczek, ``{When to Collect
  What? Optimizing Data Load via Process-driven Data Collection},'' in
  \emph{Proceedings of the 22nd International Conference on Enterprise
  Information Systems (ICEIS '20)}.\hskip 1em plus 0.5em minus 0.4em\relax
  SCITEPRESS, 2020, pp. 220--225.

\bibitem{Lippetal2021LISSU:}
J.~Lipp, S.~Sakik, M.~Kr{\"o}ger, and S.~Decker, ``{LISSU: Integrating Semantic
  Web Concepts into SOA Frameworks},'' \emph{Proceedings of the 23rd
  International Conference on Enterprise Information Systems (ICEIS '21)}, pp.
  855--865, 2021.

\bibitem{Longstaffetal2010Barriers}
T.~Longstaff, D.~Balenson, and M.~Matties, ``{Barriers to science in
  security},'' in \emph{Proceedings of the 26th Annual Computer Security
  Applications Conference (ACSAC '10)}.\hskip 1em plus 0.5em minus 0.4em\relax
  ACM, 2010, pp. 127--129.

\bibitem{Mandollaetal2019Building}
C.~Mandolla, A.~M. Petruzzelli, G.~Percoco, and A.~Urbinati, ``{Building a
  digital twin for additive manufacturing through the exploitation of
  blockchain: A case analysis of the aircraft industry},'' \emph{Computers in
  Industry}, vol. 109, pp. 134--152, 2019.

\bibitem{Mangeletal2021Data}
S.~Mangel, L.~Gleim, J.~Pennekamp, K.~Wehrle, and S.~Decker, ``{Data
  Reliability and Trustworthiness through Digital Transmission Contracts},'' in
  \emph{Proceedings of the 18th Extended Semantic Web Conference (ESWC '21)},
  vol. 12731.\hskip 1em plus 0.5em minus 0.4em\relax Springer, 2021, pp.
  265--283.

\bibitem{Mannetal2020Study}
S.~Mann, R.~Glebke, I.~Kunze, D.~Scheurenberg, R.~Sharma, U.~Reisgen,
  K.~Wehrle, and D.~Abel, ``{Study on weld seam geometry control for connected
  gas metal arc welding systems},'' in \emph{Proceedings of the 2020 17th
  International Conference on Ubiquitous Robots (UR '20)}.\hskip 1em plus 0.5em
  minus 0.4em\relax IEEE, 2020, pp. 373--379.

\bibitem{Mannetal2020Connected}
S.~Mann, J.~Pennekamp, T.~Brockhoff, A.~Farhang \emph{et~al.}, ``{Connected,
  digitalized welding production --- Secure, ubiquitous utilization of data
  across process layers},'' \emph{Advanced Joining Processes}, vol. 125, pp.
  101--118, 2020.

\bibitem{Mathuretal2016SWaT:}
A.~P. Mathur and N.~O. Tippenhauer, ``{SWaT: A Water Treatment Testbed for
  Researchand Training on ICS Security},'' in \emph{Proceedings of the 2016
  International Workshop on Cyber-physical Systems for Smart Water Networks
  (CySWater '16)}.\hskip 1em plus 0.5em minus 0.4em\relax IEEE, 2016, pp.
  31--36.

\bibitem{Maughanetal2013Crossing}
D.~Maughan, D.~Balenson, U.~Lindqvist, and Z.~Tudor, ``{Crossing the ``Valley
  of Death'': Transitioning Cybersecurity Research into Practice},'' \emph{IEEE
  Security {\&} Privacy}, vol.~11, no.~2, pp. 14--23, 2013.

\bibitem{McLaughlinetal2016The}
S.~McLaughlin, C.~Konstantinou, X.~Wang, L.~Davi, A.-R. Sadeghi, M.~Maniatakos,
  and R.~Karri, ``{The Cybersecurity Landscape in Industrial Control
  Systems},'' \emph{Proceedings of the IEEE}, vol. 104, no.~5, pp. 1039--1057,
  2016.

\bibitem{Mertensetal2021Human}
A.~Mertens, S.~P{\"u}tz, P.~Brauner, F.~Brillowski \emph{et~al.}, ``{Human
  Digital Shadow: Data-based Modeling of Users and Usage in the Internet of
  Production},'' in \emph{Proceedings of the 2021 14th International Conference
  on Human System Interaction (HSI '21)}.\hskip 1em plus 0.5em minus
  0.4em\relax IEEE, 2021.

\bibitem{Michener2015Ten}
W.~K. Michener, ``{Ten Simple Rules for Creating a Good Data Management
  Plan},'' \emph{PLOS Computational Biology}, vol.~11, no.~10, 2015.

\bibitem{Niemietzetal2020Stamping}
P.~Niemietz, J.~Pennekamp, I.~Kunze, D.~Trauth, K.~Wehrle, and T.~Bergs,
  ``{Stamping Process Modelling in an Internet of Production},'' \emph{Procedia
  Manufacturing}, vol.~49, pp. 61--68, 2020.

\bibitem{Pennekampetal2020Secure}
J.~Pennekamp, F.~Alder, R.~Matzutt, J.~T. M{\"u}hlberg, F.~Piessens, and
  K.~Wehrle, ``{Secure End-to-End Sensing in Supply Chains},'' in
  \emph{Proceedings of the 2020 IEEE Conference on Communications and Network
  Security (CNS '20)}.\hskip 1em plus 0.5em minus 0.4em\relax IEEE, 2020.

\bibitem{Pennekampetal2020Private}
J.~Pennekamp, L.~Bader, R.~Matzutt, P.~Niemietz, D.~Trauth, M.~Henze, T.~Bergs,
  and K.~Wehrle, ``{Private Multi-Hop Accountability for Supply Chains},'' in
  \emph{Proceedings of the 2020 IEEE International Conference on Communications
  Workshops (ICC Workshops '20)}.\hskip 1em plus 0.5em minus 0.4em\relax IEEE,
  2020.

\bibitem{Pennekampetal2020Privacy-Preserving}
J.~Pennekamp, E.~Buchholz, Y.~Lockner, M.~Dahlmanns \emph{et~al.},
  ``{Privacy-Preserving Production Process Parameter Exchange},'' in
  \emph{Proceedings of the 36th Annual Computer Security Applications
  Conference (ACSAC '20)}.\hskip 1em plus 0.5em minus 0.4em\relax ACM, 2020,
  pp. 510--525.

\bibitem{Pennekampetal2019Security}
J.~Pennekamp, M.~Dahlmanns, L.~Gleim, S.~Decker, and K.~Wehrle, ``{Security
  Considerations for Collaborations in an Industrial IoT-based Lab of Labs},''
  in \emph{Proceedings of the 3rd IEEE Global Conference on Internet of Things
  (GCIoT '19)}.\hskip 1em plus 0.5em minus 0.4em\relax IEEE, 2019.

\bibitem{Pennekampetal2021Confidential}
J.~Pennekamp, F.~Fuhrmann, M.~Dahlmanns, T.~Heutmann \emph{et~al.},
  ``{Confidential Computing-Induced Privacy Benefits for the Bootstrapping of
  New Business Relationships},'' RWTH Aachen University, Tech. Rep.
  RWTH-2021-09499, 2021, blitz Talk at the 2021 Cloud Computing Security
  Workshop (CCSW '21).

\bibitem{Pennekampetal2019Towards}
J.~Pennekamp, R.~Glebke, M.~Henze, T.~Meisen \emph{et~al.}, ``{Towards an
  Infrastructure Enabling the Internet of Production},'' in \emph{Proceedings
  of the 2019 IEEE International Conference on Industrial Cyber Physical
  Systems (ICPS '19)}.\hskip 1em plus 0.5em minus 0.4em\relax IEEE, 2019, pp.
  31--37.

\bibitem{Pennekampetal2019Dataflow}
J.~Pennekamp, M.~Henze, S.~Schmidt, P.~Niemietz \emph{et~al.}, ``{Dataflow
  Challenges in an \emph{Internet} of Production: A Security {\&} Privacy
  Perspective},'' in \emph{Proceedings of the ACM Workshop on Cyber-Physical
  Systems Security {\&} Privacy (CPS-SPC '19)}.\hskip 1em plus 0.5em minus
  0.4em\relax ACM, 2019, pp. 27--38.

\bibitem{Pennekampetal2021Unlocking}
J.~Pennekamp, M.~Henze, and K.~Wehrle, ``{Unlocking Secure Industrial
  Collaborations through Privacy-Preserving Computation},'' \emph{ERCIM News},
  vol. 126, pp. 24--25, 2021.

\bibitem{Pennekampetal2021The}
J.~Pennekamp, R.~Matzutt, S.~S. Kanhere, J.~Hiller, and K.~Wehrle, ``{The Road
  to Accountable and Dependable Manufacturing},'' \emph{Automation}, vol.~2,
  no.~3, pp. 202--219, 2021.

\bibitem{Pennekampetal2020Revisiting}
J.~Pennekamp, P.~Sapel, I.~B. Fink, S.~Wagner, S.~Reuter, C.~Hopmann,
  K.~Wehrle, and M.~Henze, ``{Revisiting the Privacy Needs of Real-World
  Applicable Company Benchmarking},'' in \emph{Proceedings of the 8th Workshop
  on Encrypted Computing {\&} Applied Homomorphic Cryptography (WAHC
  '20)}.\hskip 1em plus 0.5em minus 0.4em\relax HomomorphicEncryption.org,
  2020, pp. 31--44.

\bibitem{PerFail-20222022PerFail}
{PerFail 2022}, ``{PerFail 2022 -- First International Workshop on Negative
  Results in Pervasive Computing},''
  \url{https://perfail-workshop.github.io/2022/}, 2022.

\bibitem{Roepertetal2020Assessing}
L.~Roepert, M.~Dahlmanns, I.~B. Fink, J.~Pennekamp, and M.~Henze, ``{Assessing
  the Security of OPC UA Deployments},'' in \emph{Proceedings of the 1st ITG
  Workshop on IT Security (ITSec '20)}.\hskip 1em plus 0.5em minus 0.4em\relax
  University of T{\"u}bingen, 2020.

\bibitem{RWTH-Innovation-GmbH2000Offerings}
{RWTH Innovation GmbH}, ``{Offerings for Founders},''
  \url{https://www.rwth-innovation.de/en/for-start-ups-investors/offerings-founders},
  2000.

\bibitem{Samsonovetal2021Manufacturing}
V.~Samsonov, M.~Kemmerling, M.~Paegert, D.~L{\"u}tticke, F.~Sauermann,
  A.~G{\"u}tzlaff, G.~Schuh, and T.~Meisen, ``{Manufacturing Control in Job
  Shop Environments with Reinforcement Learning},'' in \emph{Proceedings of the
  13th International Conference on Agents and Artificial Intelligence (ICAART
  '21)}.\hskip 1em plus 0.5em minus 0.4em\relax SCITEPRESS, 2021, pp. 589--597.

\bibitem{Schlegeletal2020Methodological}
P.~Schlegel, D.~Buschmann, M.~Ellerich, and R.~H. Schmitt, ``{Methodological
  Assessment of Data Suitability for Defect Prediction.}'' \emph{Quality
  Innovation Prosperity}, vol.~24, no.~2, 2020.

\bibitem{Schuhetal2021Development}
G.~Schuh, A.~G{\"u}tzlaff, M.~Schmidhuber, and J.~Maibaum, ``{Development of
  Digital Shadows for Production Control},'' in \emph{Proceedings of the 2nd
  Conference on Production Systems and Logistics (CPSL '21)}.\hskip 1em plus
  0.5em minus 0.4em\relax publish-Ing., 2021, pp. 65--74.

\bibitem{Schuhetal2017Product}
G.~Schuh, S.~Rudolf, M.~Riesener, C.~D{\"o}lle, and S.~Schloesser, ``{Product
  Production Complexity Research: Developments and Opportunities},''
  \emph{Procedia CIRP}, vol.~60, pp. 344--349, 2017.

\bibitem{SEARCCH2020Sharing}
{SEARCCH}, ``{Sharing Expertise and Artifacts for Reuse through a Cybersecurity
  Community Hub},'' \url{https://searcch.cyberexperimentation.org}, 2020.

\bibitem{Serroretal2021Challenges}
M.~Serror, S.~Hack, M.~Henze, M.~Schuba, and K.~Wehrle, ``{Challenges and
  Opportunities in Securing the Industrial Internet of Things},'' \emph{IEEE
  Transactions on Industrial Informatics}, vol.~17, no.~5, pp. 2985--2996,
  2021.

\bibitem{Serroretal2020QWIN:}
M.~Serror, E.~Wagner, R.~Glebke, and K.~Wehrle, ``{QWIN: Facilitating QoS in
  Wireless Industrial Networks Through Cooperation},'' in \emph{Proceedings of
  the 19th IFIP Networking 2020 Conference (NETWORKING '20)}.\hskip 1em plus
  0.5em minus 0.4em\relax IEEE, 2020, pp. 386--394.

\bibitem{Singapore-University-of-Technology-and-Design2020iTrust}
{Singapore University of Technology and Design}, ``{iTrust - Centre for Cyber
  Security Research},'' \url{https://itrust.sutd.edu.sg/}, 2020.

\bibitem{Kouweetal2018Benchmarking}
E.~van~der Kouwe, D.~Andriesse, H.~Bos, C.~Giuffrida, and G.~Heiser,
  ``{Benchmarking Crimes: An Emerging Threat in Systems Security},''
  arXiv:1801.02381, 2018.

\bibitem{Wolsingetal2020Poster:}
K.~Wolsing, E.~Wagner, and M.~Henze, ``{Poster: Facilitating
  Protocol-independent Industrial Intrusion Detection Systems},'' in
  \emph{Proceedings of the 2020 ACM SIGSAC Conference on Computer and
  Communications Security (CCS '20)}.\hskip 1em plus 0.5em minus 0.4em\relax
  ACM, 2020, pp. 2105--2107.

\bibitem{Wolsingetal2021IPAL:}
K.~Wolsing, E.~Wagner, A.~Saillard, and M.~Henze, ``{IPAL: Breaking up Silos of
  Protocol-dependent and Domain-specific Industrial Intrusion Detection
  Systems},'' arXiv:2111.03438, 2021.

\bibitem{Xietal2021Tool}
T.~Xi, I.~M. Beninc{\'a}, S.~Kehne, M.~Fey, and C.~Brecher, ``{Tool wear
  monitoring in roughing and finishing processes based on machine internal
  data},'' \emph{The International Journal of Advanced Manufacturing
  Technology}, vol. 113, no.~11, pp. 3543--3554, 2021.

\bibitem{Xietal2020Virtual}
T.~Xi, S.~Kehne, M.~Fey, and C.~Brecher, ``{Virtual quality inspection and tool
  wear estimation based on machine internal data},'' in \emph{Proceedings of
  the 18th International Conference on Precision Engineering (ICPE '20)}.\hskip
  1em plus 0.5em minus 0.4em\relax The Japan Society for Precision Engineering,
  2020.

\bibitem{Xu2012From}
X.~Xu, ``{From cloud computing to cloud manufacturing},'' \emph{Robotics and
  Computer-Integrated Manufacturing}, vol.~28, no.~1, pp. 75--86, 2012.

\bibitem{Zhengetal2018Cybersecurity}
M.~Zheng, H.~Robbins, Z.~Chai, P.~Thapa, and T.~Moore, ``{Cybersecurity
  Research Datasets: Taxonomy and Empirical Analysis},'' in \emph{Proceedings
  of the 11th USENIX Workshop on Cyber Security Experimentation and Test (CSET
  '18)}.\hskip 1em plus 0.5em minus 0.4em\relax USENIX Association, 2018.

\bibitem{Ziegeldorfetal2017BLOOM:}
J.~H. Ziegeldorf, J.~Pennekamp, D.~Hellmanns, F.~Schwinger \emph{et~al.},
  ``{BLOOM: BLoom filter based Oblivious Outsourced Matchings},'' \emph{BMC
  Medical Genomics}, vol. 10 (Suppl 2), 2017.

\bibitem{Zilberman2020An}
N.~Zilberman, ``{An Artifact Evaluation of NDP},'' \emph{ACM SIGCOMM Computer
  Communication Review}, vol.~50, no.~2, pp. 32--36, 2020.

\bibitem{Zilbermanetal2020Thoughts}
N.~Zilberman and A.~W. Moore, ``{Thoughts about Artifact Badging},'' \emph{ACM
  SIGCOMM Computer Communication Review}, vol.~50, no.~2, pp. 60--63, 2020.

\end{thebibliography}




\end{document}